\documentclass[english,12pt, draftclsnofoot, onecolumn]{IEEEtran}
\usepackage[T1]{fontenc}
\usepackage[latin9]{inputenc}
\usepackage{float}
\usepackage{textcomp}
\usepackage{amsmath}
\usepackage{amsthm}
\usepackage{amssymb}
\usepackage{stackrel}
\usepackage{graphicx}
\usepackage{setspace}
\UseRawInputEncoding

\makeatletter

\providecommand{\tabularnewline}{\\}
\floatstyle{ruled}
\newfloat{algorithm}{tbp}{loa}
\providecommand{\algorithmname}{Algorithm}
\floatname{algorithm}{\protect\algorithmname}

\theoremstyle{plain}
\newtheorem{thm}{\protect\theoremname}
\theoremstyle{definition}
\newtheorem{defn}[thm]{\protect\definitionname}
\theoremstyle{definition}
\newtheorem{example}[thm]{\protect\examplename}
\theoremstyle{plain}
\newtheorem{cor}[thm]{\protect\corollaryname}

\usepackage{tikz}
\usetikzlibrary{positioning}
\usepackage{pgfplots}
\newdimen\R % radius
\newdimen\SmallR 
\@ifundefined{showcaptionsetup}{}{%
 \PassOptionsToPackage{caption=false}{subfig}}
\usepackage{subfig}
\makeatother

\usepackage{babel}
\usepackage{listings}
\providecommand{\corollaryname}{Corollary}
\providecommand{\definitionname}{Definition}
\providecommand{\examplename}{Example}
\providecommand{\theoremname}{Theorem}

\begin{document}

\title{Dynamic Network-Code Design for Satellite Networks}

\author{I. Shrem, B. Grinboim, and O. Amrani}
\maketitle
\begin{abstract}
\begin{singlespace}
Internet access from space enjoys renaissance as satellites in Mega-Constellations
is no longer fictitious. Network capacity, subject to power and computational
complexity constraints among other challenges, is a major goal in
this type of networks. This work studies Network Coding in the presence
of dynamically changing network conditions. The notion of \emph{generalized
acyclic network} is introduced and employed for promoting the generation
of linear-multicast network code for what is considered to be a cyclic
network. The performance of several network coding schemes, among
these is the known \emph{static network code}, is evaluated by a STK
simulation for a swarm of communicating satellites, conceptually based
on the Iridium system. Exploiting the prior knowledge of the networks
topology over time, new network coding approaches are described, whose
aim is to better cope with the time-varying, dynamic behavior of the
network. It is demonstrated that in all cases, pertaining to our example
network, static network codes under-perform compared to the presented
approach. In addition, an efficient test for identifying the most appropriate
coding approach is presented.
\end{singlespace}

\end{abstract}

\section{Introduction}

\begin{doublespace}
Satellite communication networks, and in particular integrated information
networks between space and earth, are increasingly deployed to serve
in various aspects of life. Until recently, most of the communications
among these satellites have been predominantly based on relaying via
terrestrial base stations. Ignoring the potential resource, namely
the inter satellite links (ISL), results in an atrophied network.
Such satellite networks can not meet today's growing wireless communication
needs. One of the canonical examples of this kind, relates to the
idea of providing Internet services from space using Low-Earth Orbit
(LEO) satellites, an idea that had regained popularity in recent years
\cite{PortilloOctober2018}. This renaissance may be attributed to
developments in technology and sophisticated digital communication
schemes, combined with the reduced costs of satellite development
and launching costs known to fuel the ``New-Space'' revolution.
As always, power consumption and communication throughput is a main
concern \cite{Radhakrishnan2016},\cite{MuriApr2012}.
\end{doublespace}

Network coding is a research field targeting the improvement of communications
within a network. The fundamental idea, discussed in \cite{Alshwede},
is that intermediate nodes in the network serve as computational units
rather than merely r
elays, thus promoting increased overall network
throughput. Network coding has been shown to potentially increase
overall network throughput in single-source multiple-sink communications
scenarios (i.e., broadcast of massages), or multi-source networks\cite{LiFeb2003,Yeung2006,Gamal2011}.
One of the fundamental results in this field is that in the case of
a single-source, acyclic, and finite network, given that all computations
are done over a field large enough, a \emph{linear network code} (LNC)
can achieve the maximum capacity to every \emph{eligible sink} \cite{LiFeb2003,KoetterOct2003}.
By eligible sink, we mean a sink whose max-flow (achievable throughput)
is greater or equal to the source transmission-rate.

\begin{doublespace}
A simple example demonstrating the strength of network coding is the
so-called butterfly network, as demonstrated in Figures \ref{fig:Butterfly-network-without}
and \ref{fig:Butterfly-network-with}. Therein, two signals, $a$
and $b$, are transmitted towards two sinks, $6$ and $7$. In Figure
\ref{fig:Butterfly-network-without}, the network does not employ
network coding (i.e., intermediate nodes do not execute any coding-related
computations) - hence only one of the sinks, either $6$ or $7,$
can decode both messages in each network use. In Figure \ref{fig:Butterfly-network-with},
the network employs network coding. To this end, node $4$ performs
bit-wise XOR (exclusive-or) on both its inputs, which in turn enables
both sinks to decode the two transmitted messages.

\begin{figure}
\begin{minipage}[c]{0.45\columnwidth}%
\begin{center}
\caption{\label{fig:Butterfly-network-without}Butterfly network without network
coding}
\begin{tikzpicture}[ 
regularnode/.style={circle, draw=black, fill=white, thick, minimum size=7mm}, 
locator/.style={circle, draw=white, fill=white,thin, minimum size=1mm},
node distance=0.5cm
] 
%Nodes 
\node[regularnode] (1) {1};
\node[locator] (under1) [below=of 1] {};
\node[regularnode] (2) [right=of under1] {3};
\node[regularnode] (3) [left=of under1]  {2};
\node[locator] (under3) [below=of 3] {};
\node[regularnode] (4) [right=of under3]  {4};
\node[regularnode] (5) [below=of 4]  {5};
\node[locator] (under5) [below=of 5] {};
\node[regularnode] (6) [left=of under5]  {6};
\node[regularnode] (7) [right=of under5]  {7};
  %Lines
\draw[->] (1.south east) -- (2.north west) node[near start,right] {b};
\draw[->] (1.south west) -- (3.north east) node[near start,left] {a}; 
\draw[->] (2.south west) -- (4.north east)
node[near start,left] {b};
\draw[->] (3.south east) -- (4.north west) node[near start,right] {a};
\draw[->] (4.south) -- (5.north) node[midway,right] {a/b}; 
\draw[->] (5.south east) -- (7.north west) node[near start,right] {a/b}; 
\draw[->] (5.south west) -- (6.north east) node[near start,left] {a/b};
\draw[->] (2.south east) -- (7.north east) node[midway,right] {b};
\draw[->] (3.south west) -- (6.north west) node[midway,left] {a};
\end{tikzpicture} 
\par\end{center}%
\end{minipage}\hfill{}%
\begin{minipage}[c]{0.45\columnwidth}%
\begin{center}
\caption{\label{fig:Butterfly-network-with}Butterfly network with network
coding}
\begin{tikzpicture}[ 
regularnode/.style={circle, draw=black, fill=white, thick, minimum size=7mm}, 
locator/.style={circle, draw=white, fill=white,thin, minimum size=1mm},
node distance=0.5cm
] 
%Nodes 
\node[regularnode] (1) {1};
\node[locator] (under1) [below=of 1] {};
\node[regularnode] (2) [right=of under1] {3};
\node[regularnode] (3) [left=of under1]  {2};
\node[locator] (under3) [below=of 3] {};
\node[regularnode] (4) [right=of under3]  {4};
\node[regularnode] (5) [below=of 4]  {5};
\node[locator] (under5) [below=of 5] {};
\node[regularnode] (6) [left=of under5]  {6};
\node[regularnode] (7) [right=of under5]  {7};
  %Lines
\draw[->] (1.south east) -- (2.north west) node[near start,right] {b};
\draw[->] (1.south west) -- (3.north east) node[near start,left] {a}; 
\draw[->] (2.south west) -- (4.north east)
node[near start,left] {b};
\draw[->] (3.south east) -- (4.north west) node[near start,right] {a};
\draw[->] (4.south) -- (5.north) node[midway,right] {a+b}; 
\draw[->] (5.south east) -- (7.north west) node[near start,right] {a+b}; 
\draw[->] (5.south west) -- (6.north east) node[near start,left] {a+b};
\draw[->] (2.south east) -- (7.north east) node[midway,right] {b};
\draw[->] (3.south west) -- (6.north west) node[midway,left] {a};
\end{tikzpicture} 
\par\end{center}%
\end{minipage}
\end{figure}

\end{doublespace}

Through the years, significant research efforts have been devoted
to the subject of linear network codes for acyclic and finite networks
with a single communication source. Common classification of such
networks is to linear multicast, broadcast, dispersion and generic.

\begin{doublespace}
Additional researched classes of LNC include \emph{static LNC} and
\emph{variable-rate LNC}. With respect to static network codes, given
a set of diverse network configurations (typically relating to commonly
expected link failures), the eligible sinks in any of these configurations
can still recover the source messages \cite{KoetterOct2003,Fong2006Oct2006},
without having to change the network code whenever the network configuration
changes. In variable-rate network codes, the source can generate messages
at various different rates, again, without having to modify the network
code itself \cite{Fong2006Oct2006}.

\end{doublespace}

Facilitating network coding in nano-satellite-based networks is particularly
desirable as it allows the overall network capacity to improve without
having to increase the power consumption - a scarce resource when
it comes to nano-satellites. This may be possible, since the end-to-end
topology of the network is known to the network planner and hence
accessible to the operator. Moreover, since the satellite-network
dynamics is typically known (based on Kepler's laws), \textbf{future}
topology of the network is also predictable. Note that we assume that
the topology and dynamics are only known to the operator. On the other
hand, the satellite network dynamics poses a challenge to the classical
network-coding approach, which has to be carefully tailored.

To address this challenge and provide practical tools and recommendations
to network planners, we introduce a network coding implementation
for a sample satellite network. Our network is conceptually based
on the Iridium system - one of the most prominent test-cases of a
satellite communication network \cite{Rodriguez20162016,Pratt1999}.
While Iridium is not planned as a network of small satellites, the
results and methods presented herein are readily applicable to any
satellite network (our sample network is based on Iridium in most
of the satellites parameters).

The rest of the paper is organized as follows. Section 2 describes
the network modeling and simulation environment. The following sections
provide a step-by-step implementation of the proposed LNC for an Iridium-based
network. Section 3 describes the first step of the LNC construction:
path finding. Section 4 details the basic LNC algorithm construction.
In section 5, a few different approaches for dealing with the network's
dynamics are considered and compared. Section 6 concludes the paper.

\section{Modeling the Network}

Global satellite networks have been around since the 1990's to provide
world-wide communications; among these are the Iridium and GlobStar-Immersat.
Notably, the satellite itself, being one of the main factors in the
overall cost of a satellite network, has a major impact on the service
charges.

In recent years, nano-satellite networks are planned with the aim
of significantly reducing overall costs, allowing this type of enterprises
to be lucrative. Nano-satellites, a.k.a. ``CubeSats'', are very
small ($1U=10cm\times10cm\times10cm$), and weigh around $1kg$, considerably
reducing launching costs, and allowing dozens of satellites to be
launched into space together, on a single launcher. In addition, since
nominal orbital altitudes of nano-satellites are much lower than existing
satellite networks, its components, including some of the communication
ingredients, does not necessitate ``space grade'' qualification.

In this section we shall introduce the common methodologies used to
simulate and research cubeSat networks. In addition, we shall present
the model built specifically for testing and challenging the concept
of implementing a network code over a satellite based network. 

\subsection{Satellite Constellations \& Simulations}

A trade-off to consider when planning a satellite network is coverage
vs. costs. As the launching constitutes a main cost factor, reducing
the number of launches, and hence also the number of constellation
orbital planes, is a common tactic. The main downside of such tactic
is the relatively sparse coverage it provides. A possible solution
is to create a cluster of satellites that only requires a small number
of planes with a specific zone covered. This method is also considered
for deep-space missions.

Superior coverage can be obtained by grouping satellites in different
planes. In the industry, the first planned constellation was Iridium
which initially targeted 7 planes and 77 satellites; eventually, 6
planes with 66 satellites was sufficient for providing the needed
coverage. The main shortcoming of employing CubeSats for providing
coverage is the orbital radius.Since the power available in a CubeSat
is considerably smaller than in Iridium satellite, link-budget considerations
necessitate launching CubeSats to lower orbital radius. This, in-turn,
means smaller coverage per CubeSat, so that additional orbital planes
are necessary to achieve a desired coverage.

\subsection{Link Analysis}

In general, there are 3 principle link types in a CubeSat network
(or any other satellite network for that matter) : \emph{uplink} for
transmitting commands from the ground station (GS) to the satellite,
\emph{downlink} for transmitting data to the GS, and \emph{ISL} for
transmitting data among satellites. The \emph{propagation path loss},
depends mainly on the distance $r$ between the communicating nodes,
and the frequency used for the transmission.

\subsubsection{Link Budget}

To lay down the conventions and assumptions for\emph{ }link budget
calculations of a cubeSats network, we reviewed existing and future-planned
cubeSat network enterprises (see e.g. \cite{Popescu2016March2016,Radhakrishnan2016}).
Our findings are summarized below.
\begin{enumerate}
\item \textbf{Distance} - The distance $r$ between the different transceivers
(CubeSat - GS, CubeSat - CubeSat) is determined by the orbital parameters
of the CubeSat's constellation.
\begin{enumerate}
\item The distance between the ground station and a CubeSat is determined
by the orbit trajectory pole and effective horizon; nominally in CubeSats
constellations $100_{km}\leq r_{GS-Sat}\leq2000_{km}$.
\item The distance between CubeSats varies quite considerably for different
constellations (inter-plane and intra-plane links) : $10_{km}\leq r_{ISL}\leq1000_{km}.$
\end{enumerate}
\item \textbf{Frequency} - CubeSats typically use frequencies in the VHF
and UHF bands for earth-satellites communications due to 2 main reasons:
\begin{enumerate}
\item Use of unlicensed amateur radio frequencies in the VHF bands (144
MHz to 148 MHz) and UHF bands (420 MHz to 450 MHz).
\item Relatively small atmospheric attenuation which benefits the link budget.
\end{enumerate}
CubeSat planners usually use L or S band for ISL communications. This
is due to the relatively small path loss, and because the corresponding
wavelength is short enough so that building a directed patch antenna
with a gain of $5-6dBi$ is feasible.
\item \textbf{SNR - }For digital modulation schemes, the SNR at the receiver
is given by the ratio $SNR=E_{b}/N_{0}$, where $E_{b}$, the energy
per bit, is given by $E_{b}=P_{r}/R$ with $P_{r}$ being the received
Power, and \textbf{$R$} the data rate in bits per second {[}bps{]}.
The noise spectral density $N_{0}$ is expressed by $N_{0}=k_{b}T_{s}$
where $T_{s}$ is the system noise temperature, and $k_{b}$ is Boltzmann\textquoteright s
constant. Determining the system temperature is not a trivial task
since CubeSats are not radiation protected, and the systems temperature
can vary between $100\leq T_{s}\leq1000[^{\circ}K]$. The model used
for calculating path loses in such links is the \emph{free space loss
}(FSL) $L_{p}$. It is given by $L_{p}=(\frac{4\pi df}{c})^{2}$,
where $d$ denotes the distance between the transceivers, $f$ the
operating frequency and $c$ the speed of light. The SNR is then given
by: 
\begin{equation}
SNR=E_{b}/N_{0}=\frac{P_{t}G_{t}G_{r}}{k_{b}T_{s}RL_{p}}\label{eq:SNR}
\end{equation}
where $G_{t},G_{r}$ denote the transmission and reception antenna
gains, respectively. It is common to state the equation in the logarithmic
domain: $SNR=P_{t}+G_{t}+G_{r}-L_{p}-10\log_{10}k_{b}-10\log_{10}T-10\log_{10}R$
, where $P_{t}\text{,}G_{t},G_{r},L_{p}$ are given in dB.
\end{enumerate}

\subsubsection{Transceiver \& Front End Section}

Communicating data between CubeSats requires a simple and power-efficient
baseband signal processing scheme. \cite{Popescu2016March2016,Radhakrishnan2016,MuriApr2012}
present the properties of the communication channel alongside the
requirements, limitations, and today's common practice among nano
satellites. A digital modulation method suitable for CubeSat missions
is frequency shift keying (FSK) or binary phase shift keying (BPSK),
which has been used with other low-power low-data rate applications.
Notably, the bit error probability associated with such modulation
techniques is relatively small even in low SNRs typical for these
types of links. In addition, BPSK/BFSK are very simple to implement,
hence facilitating the useof COTS (commercial off-the-shelf) components,
such as low power consuming modulators/demodulators. The main shortcoming
of this digital scheme is that the supported rate is low, hence severely
limiting the amount of data that can be sent in the short time-period
a CubeSat is visible from the ground station.

Often CubeSats do not have angular orientation control (due to limited
available power/size/weight), which necessitates the use of a wide
coverage, omni-directional or hemispherical antenna. This naturally
comes at the price of relatively low antenna gain. In addition, power
amplification is scarce due to power/weight constrains, the commonly
used transmit power, $p_{t}$, at the CubeSat side is $15_{[dBm]}\leq p_{t}\leq30_{[dBm]}$.

For the ground station the conditions are much less restrictive, typical
parameters for the antenna gain and transmission power are - $G_{GS}=15dBi$
(for a Yagi antenna), and $40_{dBm}\leq p_{t}\leq50_{dBm}$, respectively

Table \ref{table: power budget} summarizes some typical parameters:

\begin{table}[H]
\caption{Power Budget Conclusion}
\label{table: power budget}

\begin{minipage}[t]{0.45\textwidth}%
\subfloat[Inter Satellite Links]{

\begin{tabular}{|c|c|c|}
\hline 
 & L - band & S-band\tabularnewline
\hline 
\hline 
$f[GHz]$ & 1.2 & 2.4\tabularnewline
\hline 
$p_{t}[dBm]$ & 15 & 15\tabularnewline
\hline 
$G_{t}[dBi]$ & 5 & 5\tabularnewline
\hline 
$G_{r}[dBi]$ & 5 & 5\tabularnewline
\hline 
$L_{p}[dB]$ & 115 & 120\tabularnewline
\hline 
Noise Figure $N_{f}[dB]$ & 3 & 3\tabularnewline
\hline 
$T_{s}[^{\circ}K]$ & 1200 & 1200\tabularnewline
\hline 
$R[bps]$ & 3M & 1M\tabularnewline
\hline 
$SNR[dB]$ & 10 & 9\tabularnewline
\hline 
\end{tabular}}%
\end{minipage} %
\begin{minipage}[t]{0.45\columnwidth}%
\subfloat[Up/Down Links]{

\begin{tabular}{|c|c|c|}
\hline 
 & Uplink & Downlink\tabularnewline
\hline 
\hline 
$f[GHz]$ & 437 & 146\tabularnewline
\hline 
$p_{t}[dBm]$ & 15 & 50\tabularnewline
\hline 
$G_{t}[dBi]$ & 15 & 0\tabularnewline
\hline 
$G_{r}[dBi]$ & 0 & 13\tabularnewline
\hline 
$L_{p}[dB]$ & 144 & 154\tabularnewline
\hline 
Noise Figure $N_{f}[dB]$ & 2 & 5\tabularnewline
\hline 
$T_{s}[^{\circ}K]$ & 1000 & 1000\tabularnewline
\hline 
$R[bps]$ & 2400 & 1 M\tabularnewline
\hline 
$SNR[dB]$ & 12.7 & 11.35\tabularnewline
\hline 
\end{tabular}}%
\end{minipage}

\end{table}

\subsection{Modeling the Test Network}

This subsection considers the different assumptions and methods used
to generate our ``sand-box'' network. In the absence of an available
model for LEO cubeSat network, constructing a realistic model to test
the utilization of a network code in a cubeSat network is the first
step. Our model combines the topology and \emph{dynamics} of the Iridium
satellite network with the \emph{communication scheme} employed for
cubeSat networks. We capitalize on two major characteristics of the
Iridium network. First, is the network ``density'': the Iridium
network offers many potential links (especially ISLs). When examining
a network as a graph, one can see that network coding is beneficial
when the graph's max-flow is sizable (with respect to a single link).
Second, is the network dynamic behavior. Testing a network code in
a dynamic network requires paying special attention to the state of
all links at all times. We expect that the time-domain features of
the network topology will be the most significant factor when adapting
the network-code to the changing network, hence the importance of
modeling the \textbf{real} dynamics of the Iridium network.

Note that an inherent mismatch exists between the communication scheme
of a $1-10Kg$ cubeSat and $680Kg$ Iridium satellite (especially
with respect to transmission power and antenna sizes). In the following
subsections we describe the Iridium dynamics and the communication-related
adaptations made to tackle all these issues.

\subsubsection{Model Dynamics}

The Iridium network \cite{Pratt1999} is one of the most researched
satellite communication networks among GNSS, and Globstar, making
it a perfect basis for a test case, especially since there are no
full-scale, fully developed nano-satellite networks today. The Iridium
network is a LEO (low earth orbit) constellation, hence relevant to
futuristic nano-satellite constellations (unlike GNSS which is MEO).
the constellation consists of 66 satellites with $R_{earth}^{Iridium}\sim780_{km}$,
the satellites are evenly spaced in 6 different orbital planes spaced
30 degree apart, in 86 degree inclination. In addition, we assume 8 ground stations
in different locations around the globe so as to examine the earth
segment influence on the entire network (symmetry, Doppler, line of
sight, etc.). Lastly, the constellation\textquoteright s dynamic was
modeled during 24 hours using the mentioned constellation parameters
with simulation timestamp $\Delta t_{simeStep}=1_{min}$. It is important
to note that the satellite's angular dynamics was neglected .

The complete network behavior has been projected onto a 3D matrix,
$Range$, detailing all the possible links over time, as the are affected
by the constellation's dynamics and the existence of a line-of-sight
(LOS) between a pair of satellites. The value of the cell $Range_{i,j}(t)$
denots the distance between node $i$ and node $j$, $t$ minutes
from the beginning of a test scenario (When a communication link does
not exist, the value is null):

\begin{align*}
Range( & :,:,t)=\left(\begin{array}{ccc}
Range_{1,1}(t) & \cdots & Range_{1,66}(t)\\
\vdots & \ddots & \vdots\\
Range_{66,1}(t) & \cdots & Range_{66,66}(t)
\end{array}\right)
\end{align*}

\subsubsection{Model Transceiver Section}

We emphasize that many communication-related aspects, such as coding,
framing, congestion avoidance etc., are in fact transparent for the
task at hand; as network code planners -- the network code acts as
an add-on that actually affects the communicated data itself and is
independent of the connection methodology. The link parameters that
will be considered are those relevant for the calculation of the link
budget, and consequently determine the link existence.

\paragraph*{Digital modulation}

As stated above, simplicity and power efficiency are dominant factors
when choosing a digital modulation scheme for CubeSats. In order to
perform in the low SNR regime, and by that facilitate a dense network
of nodes, we have chosen BPSK modulation.

\paragraph*{Front End}

Transmission power is crucial in any communications scheme, we choose
$P_{t}$= $30dBm$, which is the most powerful amplifier the literature
mentions for cubeSats. Frequency selection for Inter-satellite links
is also quite crucial for our model. L-band and S-band are the commonly
mentioned as ISL frequencies in the literature. However, these frequencies
are chosen with respect to $d_{sat-sat}\leq2000Km$, which is not
the case with Iridium where $1000_{km}\leq d_{sat-sat}\leq6000_{km}$.\cite{Popescu2016March2016}
mentions that the link margin in the case of L-band ISL with data
rate of $R=10Kbps$, and $d_{sat-sat}\sim100Km$ is less than $10db$
(hence, not adequate for reliable communication). Thus, using the
same communications scheme in our model, with $d_{sat-sat}\sim4000Km$
is impossible (assuming data rate $R=10Kbps$). The cubeSat size and
weight prevents us from increasing the transmission power, or changing
the antenna - thus, there is no other choice but to use VHF frequency
for the ISL as well with the same omni-directional antenna. It is
important that since the Downlink and Uplink do not contribute much
to the network's density, they were not fully modeled. Specifically,
ground stations front end's are not different than satellites front
end's (in terms of power amplifiers and antenna) in our model.

\subsubsection{Link Budget}

Iridium constellation consists of 3 link types: Uplink, Downlink,
ISL. Nevertheless, not all ISL are legitimate in Iridium (even if
they meet LOS and SNR requirements). As explained, one of the main
objectives of this model, is creating as dens of a network as possible,
thus embracing VHF links as previously justified. The main influence
of this implementation (i.e using only VHF communications), is that
the dominant factor influencing the link budget is solely the range
between two nodes, disabling the traditional preference of Uplinks/Downlinks
in satellite communication (mainly thanks to high-gain ground station
antennas) causing a higher network symmetry. 

\begin{table}
\caption{Satellite communication modeled parameters (all link types)}

\centerline{

\begin{tabular}{|c|c|}
\hline 
\textbf{parameter} & \textbf{value}\tabularnewline
\hline 
\hline 
$f_{c}[MHz]$ & $146$\tabularnewline
\hline 
$R[Kbps]$ (per FDMA channel) & $6.4$\tabularnewline
\hline 
$T[^{\circ}K]$ & $1000$\tabularnewline
\hline 
$G_{r}[dBi]$ & $0$\tabularnewline
\hline 
$G_{t}[dBi]$ & $0$\tabularnewline
\hline 
$P_{t}[dBm]$ & $30$\tabularnewline
\hline 
\end{tabular}

}
\end{table}

Now, after presenting all assumptions for the communication scheme,
we can define and stock a so-called \emph{link budget matrix} and
derive the link capacities according to Shannon--Hartley's Theorem
and Equation \ref{eq:SNR}.

The resultant network matrix holds the link capacities 
\begin{equation}
C_{i,j}=BW\cdot log(1+SNR_{i,j})\label{eq:Capacity matrix}
\end{equation}
 for any link between nodes \emph{i} and \emph{j}. We note that since
the satellites antennae are omni directional, and the transmission
powers are all the same, this matrix is symmetric. From this matrix,
the transformation to a graph is simple - each satellite is a node,
and the weights\footnote{we note here that greater weights indicate better communication link.}
of the connecting edges are given by $C_{i,j}$. Null $C_{i,j}$ means
that there is no edge connecting node $i$ to node $j.$

\begin{figure}
\caption{Integer Capacity Channels Modeling}

\label{fig:multi_channel}

\centerline{

\includegraphics[height=0.1\textheight]{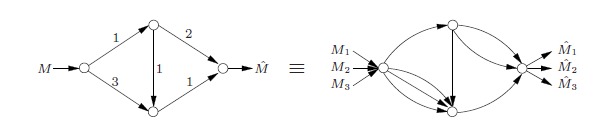}

}
\end{figure}

Finally, to be able to apply algorithms that require all edges of
a graph to be unit capacity links, we needed to manipulate our weighted
graph. To this end, the weighted graph is converted into a \emph{multi-graph}
by: 1. rounding down the capacity value of each edge, and 2. duplicating
the unit capacity edge by the resulted integer capacity, as illustrated
in Figure \ref{fig:multi_channel} \cite{Gamal2011}.

\section{Instantaneous Network Code Implementation}

The algorithm for establishing a \emph{linear network code (}LNC)
consists of several steps: 
\begin{enumerate}
\item An \emph{acyclic} graph representation of the network $G$ is given
along with defined source node and set of designated sink nodes as
introduced in the previous section;
\item The network's ($G$'s) max-flow, $h$, from the source to the designated
sinks is derived, and $h$ independent paths between the source and
each designated sink are identified;
\item The paths found in Step 2 are used: the network nodes are examined
in an upstream-to-downstream order, and for each node all its outgoing
edges are examined separately in an arbitrary order: The \emph{Global
Encoding Kernel }(GEK) for each edge is calculated so that it satisfies
the conditions in \ref{def:GEK}. This guarantees that the input to
each of the designated sinks is decodable.
\end{enumerate}
\begin{doublespace}
Steps 2 and 3 will be thoroughly explained in the Subsections 3.1
and 3.2. Subsection 3.3 introduces the concept of \emph{cyclic network
codes }(CNC) for \emph{cyclic} graphs.

A network node in $G$ that is neither source, nor sink, shall be
called an ``intermediate node''. A ``path'' is a set of links
that provides a connection between two nodes in the network.
\end{doublespace}

\subsection{LNC on acyclic graphs. }

\begin{doublespace}
Consider a network $G'=\{V,E\}$, with $V$ and $E$ denoting network
nodes and links (alternative terms for vertices and edges as we are
dealing with a communication network), respectively; a source node
$s$ in $V$ and a set of $d$ sink nodes $T=\{t_{1},\ldots t_{d}\}$
in $V$. The max-flow between $s$ and $t_{i}\in$$T$ is denoted
by $h_{t_{i}}$; in this paper, the term ``max-flow'' of a sink
$t_{i}$ shall always refer to this $h_{t_{i}}$. All the information
symbols are regarded as elements of a base field $F$ (this terminology
is often referred to as \emph{scalar network codes}).

The operation \emph{rate }of the communication network, denoted by
$r$, is a positive integer representing the number of symbols created
by the source every network use. In this paper, assume that $r$ is
constant.
\end{doublespace}

For every node $x\in V\setminus\{s\}$, denote by $In(x)$ and $Out(x)$
the sets of input and output edges to the node $x$, respectively.
For convenience, it is customary to add another node $s'$, referred
to as the \emph{imaginary source}, that has $r$ outgoing edges to
the original source $s$ - referred to as the $r$ \emph{imaginary
links} (namely, $In(s)$ represents the set of $r$ imaginary links).
Assume that $G'$ includes the imaginary source associated edges.

\begin{doublespace}
The basic concept of network coding is that all the nodes in the network
are able to perform calculations over the field $F$. 

The following is a description of a \emph{linear network code} that
is derived from \cite{Yeung2006}:
\end{doublespace}
\begin{defn}
\begin{singlespace}
\label{def:GEK} An $r$-dimensional $F$-valued linear network code
operating in an acyclic communication network is characterized by:
\end{singlespace}
\begin{enumerate}
\item \emph{Local Encoding Kernel }(LEK) - $\{k_{d,e}\}$ - A set of scalars
$k_{d,e}\in F$, one for every adjacent pair of edges $(d,e)$ in
the network, where $d\in In(x)$ and $e\in Out(x)$;
\item \emph{Global Encoding Kernel} (GEK) - $\{f_{e}\}_{e\in E}$ - A set
of $r$-dimensional column vectors $f_{e}$, one for every edge $e$
in the network such that:
\begin{enumerate}
\begin{singlespace}
\item For every non-source node $x$, and every $e\in Out(x)$, $f_{e}=\sum_{d\in In(x)}k_{d,e}f_{d}$;
\item For the imaginary links $e\in In(s)$, the set of vectors $\{f_{e}\}_{e\in In(s)}$
are defined as the $r$ linearly independent vectors that constitute
the natural basis of the vector space $F^{r}$.
\end{singlespace}
\end{enumerate}
\end{enumerate}
\begin{doublespace}
The local encoding kernel associated with node $x$ refers to a $|In(x)|\times|Out(x)|$
matrix. The vector $f_{e}$ is called the \emph{g}lobal encoding kernel
for edge $e$. 
\end{doublespace}

Note that given the local encoding kernels at all the nodes in an
acyclic network, the global encoding kernels can be calculated recursively
in an \emph{upstream-to-downstream} order based on the given definition.
\begin{defn}
\label{def:linear multicastt}Let $\{f_{e}\}$ denote the global encoding
kernels in an $r$-dimensional, $F$-valued, linear network code in
a single-source finite acyclic network. Let $V_{x}=span\{f_{d}|d\in In(x)\}$.
Then, a linear network code is a \emph{linear multicast} if $dim(V_{x})=r$
for every non-source node $x$ satisfying $max-flow(x)\geq r$.

A known algorithm for constructing a linear multicast for a single-source
finite acyclic network \cite{LiFeb2003,Yeung2006} requires the field
size $|F|$ to be greater than the size of the set of sinks (whose
$max-flow(\{T\})\geq r$) i.e, $|F|>|\{T:maxflow(\{T\})\geq r\}|$.
This algorithms provides a the LEK and GEK of a linear multicast on
the graph, i.e. a LNC that allows all target sinks to receive decodable
information simultaneously in every network use.
\begin{defn}
\label{def:upstream-to-downstream order} Let $G=(V,E)$ be a connected
directed acyclic graph, and let $s\in V$ be a node in $G$, called
the \emph{source node}. Note that $G$ is a tree, with $s$ as its
root. Denote $n=|V|$ as the number of nodes in $G$. An \emph{upstream
to downstream order} defines ordering of the nodes in $V$, $O:V\rightarrow\{1:n\}$
in $G$, such that for every $v,u\in V$, if $u$ is a parent of $v$
in the tree induced by $G$, $O(v)>O(u)$, and for every $u_{1}\not=u_{2}$,
$O(u_{1})\not=O(u_{2})$. In other words, looking on $O$ as an order
of choosing the nodes, a node is chosen only after all of its parent
nodes have been chosen. The order could also be denoted $O=\{v_{1},...,v_{n}\}$,
meaning that $O(v_{i})=i$.
\end{defn}
\end{defn}
\end{defn}
\begin{example}
Examine the graph $G$ presented in Figure \ref{fig:D2U Order Graph},
and the orders $O_{1}=\{1,2,3,4,5,6,7\}$, $O_{2}=\{1,2,4,5,3,6,7\}$,
$O_{3}=\{1,2,4,3,5,6,7\}$ and $O_{4}=\{1,3,6,7,2,4,5\}$. $O_{1}$,
$O_{2}$ and $O_{3}$ are all upstream-to-downstream orders - in all
of them, the parents of every node comes prior to them in these orders.
In contrary, $O_{4}$ in NOT an upstream-to-downstream order, since
$O_{4}(7)<O_{4}(5)$, even though $5$ is a parent of $7$ in $G$.
\begin{figure}[h]
\centering{}\caption{\label{fig:D2U Order Graph}An Example Graph $G'$}
\includegraphics[height=0.15\paperheight]{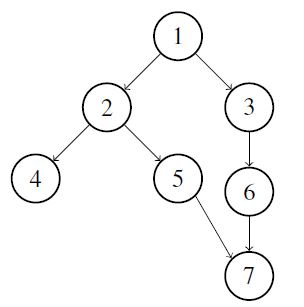}
\end{figure}
\end{example}

\subsection{Path Finding}

This subsection will describe the methodology used finding \emph{max-flow}
different paths from the source $s$ to the sinks $T$ over the graph
$G$. Path finding is needed in order to provide an ordered list that
will enable calculating in an upstream to downstream order over the
nodes as defined in Figure \ref{fig:D2U Order Graph}.

Given a graph $G$, our construction algorithm of the network code
consists of two preliminary steps; the first, calculating the \emph{max-flow}
of the graph. We used MATLAB\textquoteright s built-in function (based
on Boykov-Kolmogorov algorithm \cite{Boykov2004}) in this implementation.
The second, finding \emph{max-flow} disjoint paths from the source
$s$ to each defined sink $t\in T$. Path-finding does not necessarily
result unique paths selection for a specific configuration of source
and sinks. To dissolve this ambiguity, the parameter taken into consideration
inhere is the comparability to routing techniques. Since most routing
algorithms base on shortest-path finding (using Dijkstra's algorithm
for example), we employ the following shortest-path-based algorithm
stated for a single source \emph{$s$}, and single sink $t\in T$.

\begin{algorithm}[H]

\caption{\label{alg:shortest-path-based}shortest path based path finding}
\begin{lstlisting}
Initialization: 
Paths = []; Network = G; Weights = {C(i,j)}; source = s, sink = t;
For i = 1: max-flow
	Paths(i) = Dijkstra(Network,s,t)
// find shortest path to sink  from source  given the network Network

Weights(paths(i)) = Weights(paths(i)) -1 ; 
// remove one edge from the multi-graph

If Weights(paths(i)) =0 
	Network = Network\Paths(i); 
End
\end{lstlisting}

\end{algorithm}

Assuming a Dijkstra based routing algorithm, one could simply compare
the throughput at a sink $t$ when using routing and network coding.
However, as demonstrated in Example \ref{fig:Shortest_path_and_max_flow},
our shortest-path-based search algorithm does not guarantee finding
\emph{max-flow} disjoint paths.We did not encounter a failure finding
\emph{max-flow} paths on the Iridium network for any topology tested
(many variations over source and sinks selections). Nevertheless,
a comparison of the number of paths found by the algorithm to the
network\textquoteright s \emph{max-flow} is a required sanity check.
\begin{center}
\begin{figure}
\caption{Shortest-Path based path-finding failure. The source node $s$ is
node $\#1$, the algorithm fails to achieve 2 disjoint paths to sink
$\#10$.\label{fig:Shortest_path_and_max_flow}}

\centerline{

\includegraphics[height=0.2\textheight]{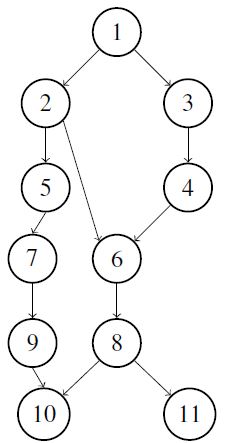}}
\end{figure}
 If the check were to fail, we would have used the Ford-Fulkerson
algorithm \cite{Johnsen1962194} (This algorithm guarantees finding
disjoint \emph{max-flow} paths). However, this algorithm does not
necessarily find the shortest path, thus disabling the opportunity
to compare our results to routing techniques.
\par\end{center}

After finding all \emph{max-flow} paths, we will observe only the
graph $G'$, which is a pruned version of the graph $G$ after removing
the edges that are not participating in any of the allocated paths.
Such a graph $G'$ will be referred to as a (pruned) paths-graph of
the network $G$ with respect to the source $s$ and set of sinks
$T$. 

\subsection{CNC on cyclic graphs. }

In the case of a finite, single-source and \emph{cyclic} network,
it has been shown \cite{Li20062006} that if the field $F$ is large
enough, there exists a Convolutional Network Code (CNC) achieving
rate of \emph{$r$} messages in each designated sink $t$. A designated
sink is one with max-flow $h_{t}\ge r;$ all the operations are done
over the field $F$, and the vectors of the GEK are vectors in $F[D]^{r}$
($D$ being a unit delay). In \cite{Erez2010}, a relatively efficient
algorithm for finding such a CNC is presented -- though it is of
much higher complexity in comparison to the efficient LNC-finding
algorithms. Consider that the higher complexity of such algorithms
derives from the fact that in the absence of an upstream-to-downstream
order on the nodes in cyclic networks, the algorithm has to visit
every node several times to make sure that the linear independency
is kept \cite{Erez2010}.

\section{Linear Multicast in Generalized Acyclic Networks}

\subsection{Definition and Motivation}

Continuing the previous section, with the aim of designing an efficient
network code for the example network with a pruned paths-graph $G'$,
it is necessary to determine whether $G'$ is a cyclic or an acylic
graph. Seemingly, it is a simple problem; the original Iridium network,
$G$, is obviously cyclic since all its channels are assumed to be
full-duplex, meaning that every link is itself a cycle. Consequently,
when considering the high connectivity and high max-flows of the example
network as modeled, it is not difficult to realize that cycles in
$G'$ are created with almost any combination of sinks. Notwithstanding
the above, since it is considerably more complex to construct a cyclic
convolutional network code than a linear multicast code for an acyclic
network, the motivation to somehow apply acyclic graph-methodologies
is high. The problem is that finding an upstream-to-downstream order,
which is an essential step of a linear multicast construction, is
of course impossible on the cyclic graph $G'$. In some cases (which
cases exactly - will be discussed later in that section), the network
operation can be characterized as a degenerate version of a cyclic
network, termed \emph{generalized acyclicity},\emph{ }which is sufficient
for applying the methodologies of acyclic graphs. This characteristic
and the relevant method to be applied is explained next.
\begin{defn}
\label{def:line graph}Assume given a graph $G=(V,E)$ (for notational
brevity, and w.l.o.g, assume that $G=G'$), a source node \emph{s},
a set of sinks $T$ and a choice of paths, denoted by $P$, from $s$
to each sink in $T$. Let the \emph{paths line-graph} (PLG) induced
from $G$, with respect to the set of paths $P$, be the graph $LG=(LV,LE)$,
with the \emph{line-graph nodes }being $LV=\{s,T,E\}$, and \emph{line-graph
edges }$(e_{1},e_{2})=\epsilon$$\in LE$ if:
\end{defn}
\begin{enumerate}
\item $e_{1}=s$, $e_{2}\in E$, where $e_{2}$ is an outgoing edge from
$s$ in $G$.
\item $e_{2}=t\in T$, $e_{1}\in E$, were $e_{1}$ is an incoming edge
to $t$ in $G$.
\item $e_{1},e_{2}\in E$ and a node $v\in V$ such that $e_{1},e_{2}$
are an incoming edge to $v$ and an outgoing edge from $v$ respectively,
and there exists a path in $P$ that includes the transition from
$e_{1}$ to $e_{2}$ via $v$.
\end{enumerate}
This type of line-graph construction was introduced by Erez \& Feder
\cite{Erez2010}. Note that a paths line-graph is not a conventional
line-graph. PLG includes line-graph edges between the edges of the
graph $G$ only if the transition between the latter edges is included
in a certain path, and only in the direction of a path. As a result,
unlike conventional line-graph that preserves the cyclic (or acyclic)
characteristics of the original graph -- a paths line-graph may be
acyclic also when the graph it is induced from is cyclic. This motivates
the following definition:
\begin{defn}
\begin{doublespace}
Given a graph $G$, a source node $s$ and a set of sinks $T$, $G$
is a\emph{ generalized acyclic network} with respect to $s$ and $T$,
if there exists a set of max-flow paths $P$ from $s$ to each sink
in $T$, so that the path line-graph induced from $G$, with respect
to $P$, is acyclic.
\end{doublespace}

Note that every acyclic graph $G$ is also generalized acyclic - since
all its PLG are necessarily acyclic. The interesting case is hence
the case in which a \emph{cyclic }graph is generalized acyclic. Example
\ref{exa:Observe-the-graph} demonstrates such a case.
\end{defn}
\begin{example}
\label{exa:Observe-the-graph}Consider the graph $G$ at the left-hand-side
of Figure \ref{Fig. line_graph_demo_8}, with the node $s=1$ as the
source and nodes $T=\{6,7\}$ as sinks. Not only $G$ is a cyclic
graph, it is also easy to see that both the edges $(4,6)$ and $(6,4)$
are necessary in order to achieve a communication rate of \emph{$max-flow=2$}
between $s$ and $T$, and hence the paths-graph $G'$ is also cyclic.
The right-hand-side figure demonstrates a PLG of $G$. In this case,
since there is only one choice of paths that achieves the max-flow
to each of the sinks, this PLG is unique. Note that even though $G'$
is cyclic, the PLG is acyclic, hence $G$ is a generalized acyclic
graph.

\begin{figure}
\caption{\label{fig:Graphical-demonstration-for}Graphical demonstration for
Example \ref{exa:Observe-the-graph}}
\label{Fig. line_graph_demo_8}

\begin{minipage}[c]{0.45\columnwidth}%
\begin{center}
\subfloat[\label{fig:generalized acyclic graph} Original graph $G$. $G$ is
cyclic, but also generalized acyclic.]{

\includegraphics[width=0.45\columnwidth]{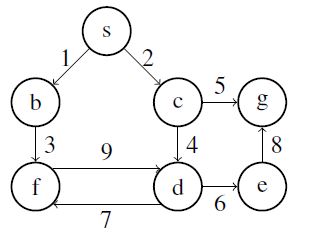}}
\par\end{center}%
\end{minipage}\hfill{}%
\begin{minipage}[c]{0.45\columnwidth}%
\begin{center}
\subfloat[\label{fig:generalized acuclic LG}$PLG$, the paths line-graph for
$G$.]{

\includegraphics[width=0.45\columnwidth]{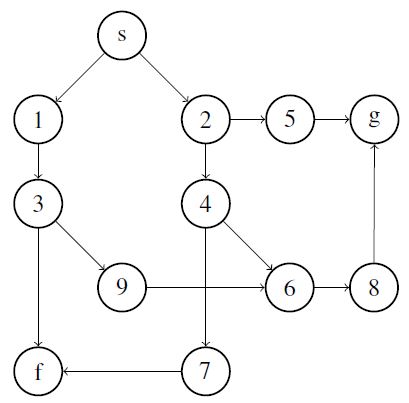}}
\par\end{center}%
\end{minipage}
\end{figure}
\end{example}
\begin{thm}
Given a generalized acyclic network $G$ with source $s$ and set
of sinks $T$, there exists a linear multicast on $G$, if the messages
and computations are carried out over a large enough field.
\end{thm}
The theorem is based on the ability to find an upstream-to-downstream
order in the paths line-graph, which essentially allows to \textquotedbl break\textquotedbl{}
cycles that exist in the original graph (but effectively not used
by any combination of paths).

A constructive proof is next given in the form of an algorithm consisting
of two stages:
\begin{enumerate}
\item Derivation of a modified version of linear multicast on the PLG;
\item Conversion of modified version to a linear multicast on $G$.
\end{enumerate}
\begin{IEEEproof}
Let $LG$ be an acyclic PLG of $G$, with $r$ being the operation
rate of the network. Since $LG$ is acyclic, it is possible to find
an upstream-to-downstream order for its nodes going from $s$ to $T$.
Note that this does not necessarily induce an upstream-to-downstream
order on the nodes of $G$. Next, the algorithm proceeds in a similar
manner to the known linear multicast algorithm (see \cite{LiFeb2003}).

For this process to yield a network code not only on the path graph
but also on the original graph $G$, every LG-\emph{node} (which is
equivalent to an edge on the graph $G$) is only allowed to be associated
with a single GEK, which is a $F^{r}$ vector. This is different from
the original algorithm in \cite{LiFeb2003}, according to which every
\emph{edge }is associated with a global encoding kernel - hence allowing
every node to ``hold'' all the global encoding kernel values of
incoming edges simultaneously. The LG-node GEK vector will be used
for determining the GEK value of the equivalent graph $G$ edge later
in this algorithm.

For example, consider LG-node $6$ in Figure \ref{fig:Graphical-demonstration-for},
which has two incoming edges. If each node is allowed to hold more
than one GEK vector, the conversion to a linear network code on $G$
will not be straight forward. This is because $6$ is an edge in the
original graph $G$, and hence allowed to hold only a single GEK value
according to definition \ref{def:GEK}.

This calls for a modification in the aforementioned algorithm (\cite{LiFeb2003}):
instead of running over the nodes and edges of the $LG$, the algorithm
has to run only on its nodes. The paths for the modified algorithm
are those induced by the paths of the original graph $G$.

Except for the source node \emph{$s$} (which is treated separately),
the modified algorithm guarantees that every node (anywhere) in the
line-graph that leads to a non-empty subset of sinks $T'\subseteq T$
holds only one vector that satisfies two conditions:
\begin{enumerate}
\item It is linearly independent from the set of all the other vectors held
by the nodes that are in paths to sinks in $T'$ (for each $t'\in T'$
separately, of course). For every sink, there are exactly $r$ such
nodes, since they are induced by edges in $r$ distinct paths to that
sink.
\item It is a linear combination of the vectors held by its incoming edges.
\end{enumerate}
The existence of such vectors is a straight-forward corollary from
the original algorithm given for linear multicast \cite{LiFeb2003}. 

Contrary to the original linear multicast algorithm (\cite{LiFeb2003}),
where each of the outgoing edges from a node may hold a different
GEK, in the proposed algorithm all the edges coming out of a node
share the same GEK; once the specific GEK vector of a node has been
determined, all its outgoing edges obtain the same GEK value. Note
that since the proposed algorithm operates in an upstream-to-downstream
order, all the GEK values of incoming edges to a node are determined
before calculating the node's GEK value.

As for the source node, the algorithm is initialized with the source
incoming edges holding all the unit vectors of the $r$ dimensional
vector space. The values of its outgoing edges are not initialized,
but rather determined based on the values of the other nodes connected
to them (which are equivalent to the outgoing edges from the source
in the original graph, so that they fulfill condition 1 above). Thus,
each sink is receiving $r$ linearly independent vectors, which means
that a valid linear multicast on the LG is obtained in accordance
with Definition \ref{exa:Observe-the-graph}.

Given the linear multicast on the LG with a single vector value for
each and every node, conversion to a linear multicast on $G$ is straight
forward: the global encoding kernel of every edge $e\in E$, its will
be the global encoding kernel of its equivalent LG-node. It follows
directly from the construction of the linear multicast on the graph
$LG$, that the result is indeed a linear multicast on $G$.
\end{IEEEproof}

\subsection{Algorithm Complexity Analysis}

As expected - the computational complexity of the proposed algorithm
is lower than the alternatives, as the latter involve the design of
a cyclic network code. In \cite{Erez2010}, the complexity of finding
a convolution network code for a cyclic network is $O(d^{3}|E|^{w+2}))$
with $d=|T|$ and $w$ is the exponent of multiplying two matrices
which is known to be $2\le w<2.37$. The efficiency of the algorithm
presented here is identical to that of a regular linear multicast
construction, which if implemented by the method shown in \cite{Jaggi2005},
is $O(|E|dh^{2}+|E|hd^{2})$ -- the path-finding algorithm needs
run only on the original graph, while the code construction is only
carried out on the PLG. The LG construction is bounded by $|E|\cdot h$. 

\section{Network Code Under Dynamic network conditions}

Up until this point the paper considered linear multicast for a snapshot,
i.e. instantaneous setting, of the satellite constellation network.
From now on, dynamic behavior of the network is considered to account
for the constantly changing topology of the network. It is clear that
applying the network code designed for a specific network topology
on another topology may render the decoding in some sinks impossible.
The network code must be properly adapted, and its parameters distributed
among the relevant network nodes, whenever the network topology changes
enough. Alternatively, one may try to come up with code constructions
that are effective even under topology changes. In order to minimize
the impact of network topology changes on communication performance,
we shall utilize the fact that these changes may have predictable
behavior. In this section, three coding methods that utilize this
network behavior are described and compared.
\begin{enumerate}
\item \textbf{Instantaneous Network Code - Construction and distribution
of a new network code with each network change.} Given that the topology
of the entire network is only known to a ground station, this method
calls for the distribution of tailored network code-parameters for
each time interval - call it a\emph{ 'Genie'} approach. The distributed
message size is $|V|^{3}\centerdot log_{2}(|T|+1)$; where $log_{2}(|T|+1)$
denotes the symbol rate, and $|V|^{3}$ amounts to choosing the triplet
$(x,d=In(x)$ , $e=Out(x)$) (per node $x$). Assuming communication
rate of $6.4Kbps$, and $66$ satellites (as presented in chapter
2), the time for distributing the new parameters throughout the network
is: 
\begin{equation}
t_{distribution}=\frac{|V|^{3}\text{\ensuremath{\centerdot}}log_{2}(|T|+1)}{6400}[sec]\approx44\cdot|F|[sec]\label{eq:t_distribution}
\end{equation}
. Given that the network topology changes every $60_{sec}$, this
approach is practically irrelevant. In this section, this method is,
nevertheless, employed as upper bound. Note that if memory resources
(at the network nodes) are abundant, and the network topology is truly
periodic, then the code parameters, on its various derivatives, can
be stored onboard and there is no need for re-distribution when the
network changes. This, however, is not the general case that we consider
in this work.
\item \textbf{Static Network Code.} \emph{Static network coding} \cite{Fong2006Oct2006,KoetterOct2003}
is an approach that enables a network code to account for different\textbf{
pre-defined} topology configurations, denoted by $\varepsilon\in\hat{\varepsilon}$.
For each configuration $\varepsilon$, a subset of edges $E_{\varepsilon}\subseteq E$
remain, and all the other edges are omitted from the network, so that
the graph is $G_{\varepsilon}=\{V,E_{\varepsilon}\}$\footnote{if one would have wanted to be prepared for a node failure, the configuration
would include the subset of edges that are not adjacent to this node.}. A static code with respect to $G$ and $\hat{\varepsilon}$, could
operate under any configuration $\varepsilon\in\hat{\varepsilon}$,
but it requires working over a field $\hat{F}$, which is growing
proportionally to the number of configurations in $\hat{\varepsilon}$
(One field element has to be added for each pair of configuration-sink
$(\varepsilon,t)$ such that under $\varepsilon$, the paths from
the source to $t$ differ from those in the other configuration) \cite{qifu2018}.
If a network behaves in a predictable manner, then the network dynamics
can be taken into account; it enables us to consider the network over
$\tau$ time intervals as $\tau$ configurations derived from a ''unified''
network, defined as $\stackrel[t=1]{\tau}{\cup}G_{\varepsilon_{t}}=\{V,\stackrel[t=1]{\tau}{\cup}E_{\varepsilon_{t}}\}$
($\varepsilon_{t}$ denoting the \emph{links-failure} configuration
during time interval $1\leq t\leq\tau$). Using the unified network
approach, and regarding all possible network topologies as links-failure
configurations, enables one to deal not only with link-failures, but
also with link additions. The main shortcoming of the static network
coding approach, as will be later shown when the methods are compared,
is the significantly larger field size it requires, which results
with a corresponding reduction in communication rate.
\item \textbf{Intersection Network Code.} Another possible approach for
coping with the network dynamics is to use only the links that exist
for all the network configurations during a time interval $\tau$.
According to this approach, the network code is designed based on
an effective network given by: $\stackrel[t=1]{\tau}{\cap}G_{\varepsilon_{t}}=\{V,\stackrel[t=1]{\tau}{\cap}E_{\varepsilon_{t}}\}$.
Using this method, the field size does not change even though the
network is dynamic - yet the communication rate is smaller as a result
of the reduction in the max-flow. The trade off in choosing $\tau$
is clear: too large $\tau$ results with a relatively small communication
rate due to low achievable max-flow (in case the network changes considerably
over time); too small $\tau$ waists channel resources by needing
to re-distribute updated code parameters.
\end{enumerate}

\subsection{Achievable Communication Rate - Upper Bound}

\begin{figure}[H]
\caption{\label{fig:optimal-rate-bound}Optimum rate upper bound from the source
to sinks \{6,13,15\}. Simulation interval: $t=0$ to $t=360$ {[}$min.]$}

\centerline{

\includegraphics[height=0.2\textheight]{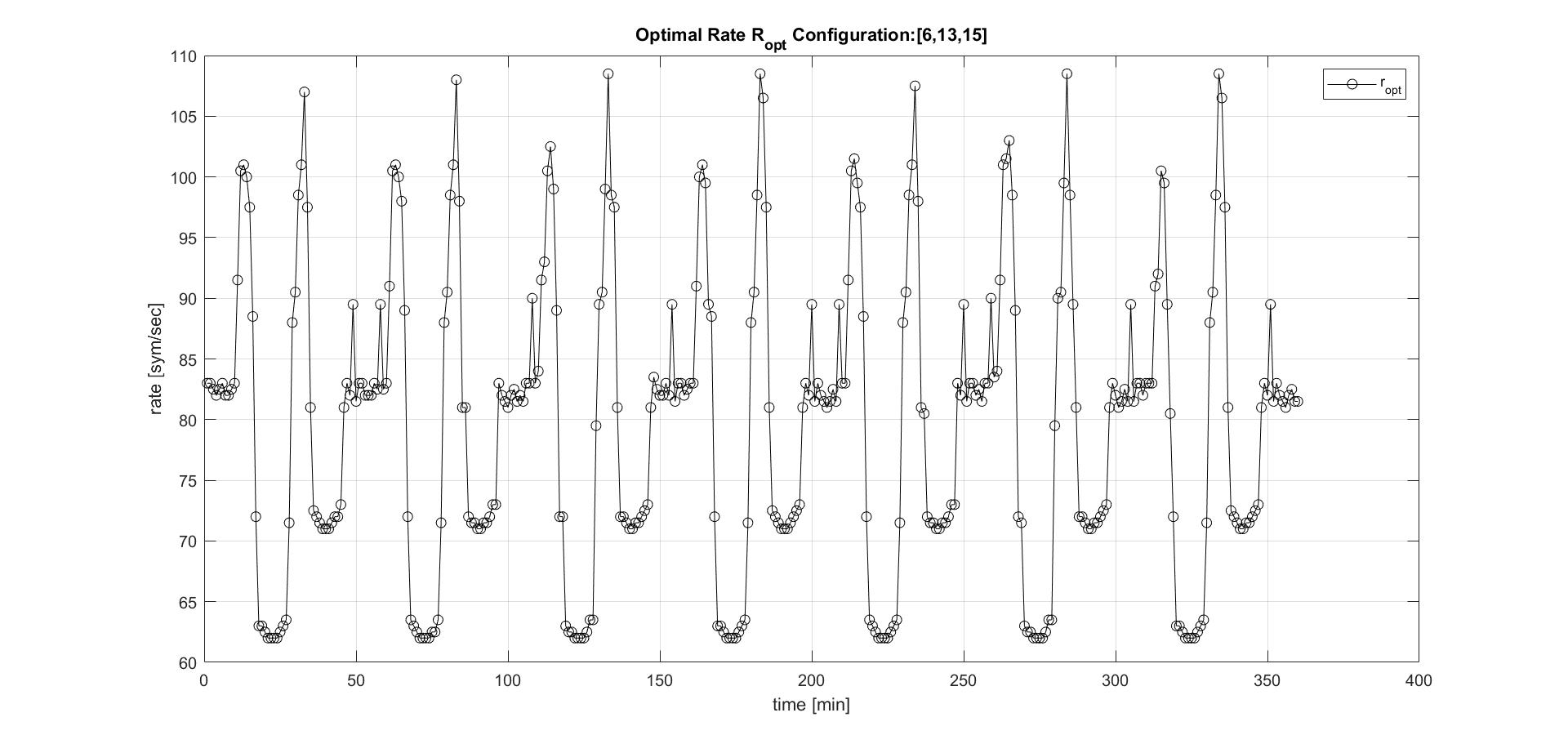}}
\end{figure}

Prior to comparing the aforementioned methods, first, an upper bound
on the communication rate $r_{opt}$, given by the instantaneous network-code
communication rate, is calculated for an example scenario and drawn
in Figure \ref{fig:optimal-rate-bound}.

The max-flow is calculated according to \cite{Boykov2004}, where
input link weights are set by eq. \ref{eq:Capacity matrix}. The overall
rate is calculated by dividing the max-flow value by $\lceil log_{2}|F|\rceil$,
the bit rate induced by the field $F$. The observed max-flow periodicity
over time is of course a consequence of the network-topology periodicity.

Note: the sinks are numbered in a sequential order, such that the
nodes in the first plane are numbered $1-11$, the second $12-23$
etc. Adjacent nodes in a plane are numbered sequentially, hence node
pairs $\{1,12\},\{2,13\}$ etc. represent the same positions within
the planes. The sinks used in the simulation shown in Figure \ref{fig:optimal-rate-bound}
are taken from two different planes $\{2,6\}$\&$\{13\}$.

\subsection{Identifying the best Coding Scheme}

Next, we proceed to compare the aforementioned coding approaches -
using the methodology of \emph{static network codes}, and \emph{intersection
network codes, }along with an introduction of \emph{network interval-rate.}

\subsubsection{\emph{Network Static and Intersection Rates}}
\begin{figure}
\caption{\label{fig:techniques-comparision.-it}Techniques comparison: left
figure - the interval rate with $T=10[min]$ outperforms the intersection
rate; right figure - the intersection rate outperforms the interval
rate for any selection of \emph{sub-interval time} $T$. }
\begin{minipage}[t]{0.45\columnwidth}%
\subfloat[{\label{interval outperforms}sinks:$T=\{6,13,15\}$ the interval rate
with $T=10[min]$}]{\includegraphics[width=3.15in]{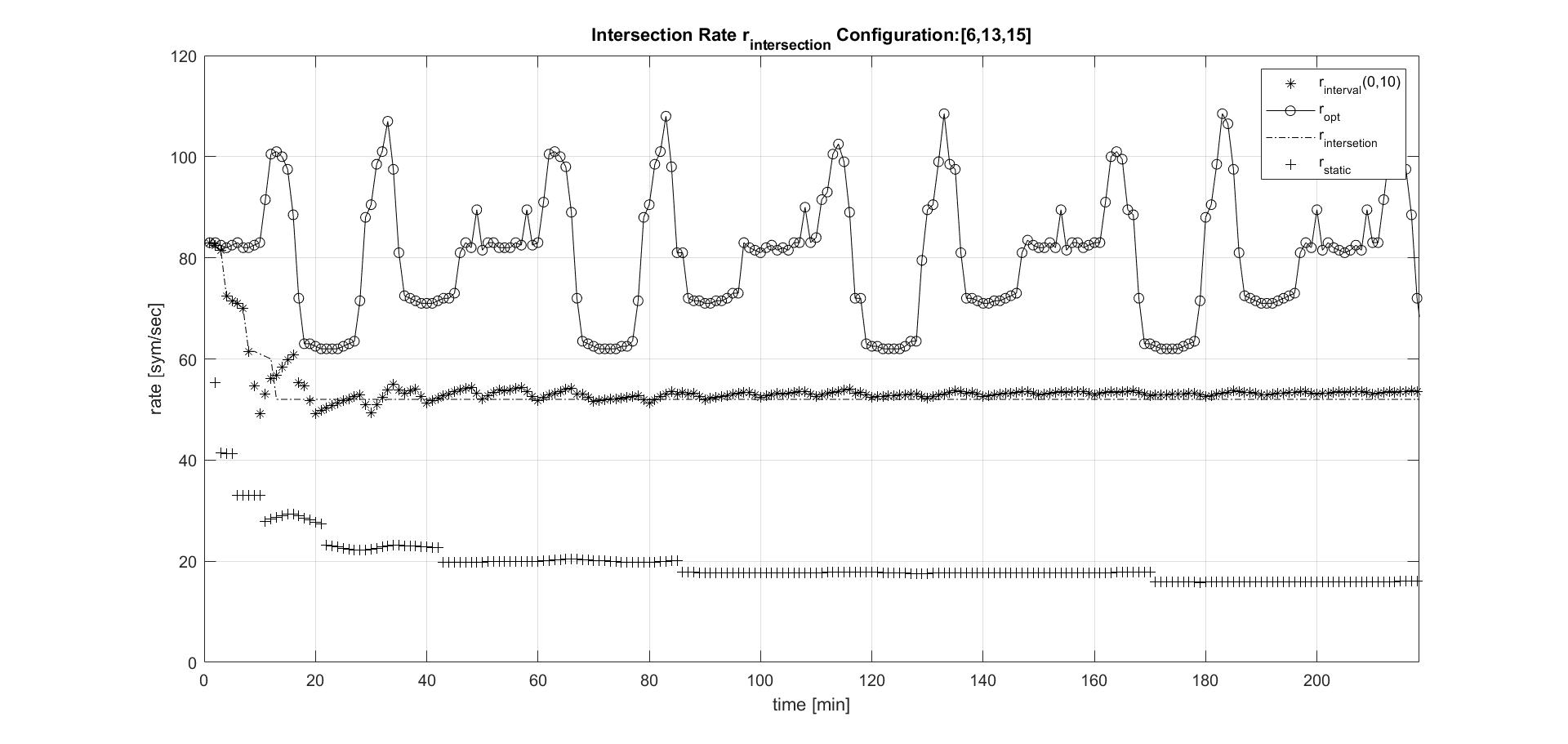}}%
\end{minipage}\hfill{}%
\begin{minipage}[t]{0.45\columnwidth}%
\subfloat[\label{fig:sinks-:-intersection outperforms}sinks:$T=\{15,17,19\}$
- the intersection rate outperforms the interval rate for any selection
of \emph{sub-interval time} $T$.]{

\includegraphics[width=3.15in]{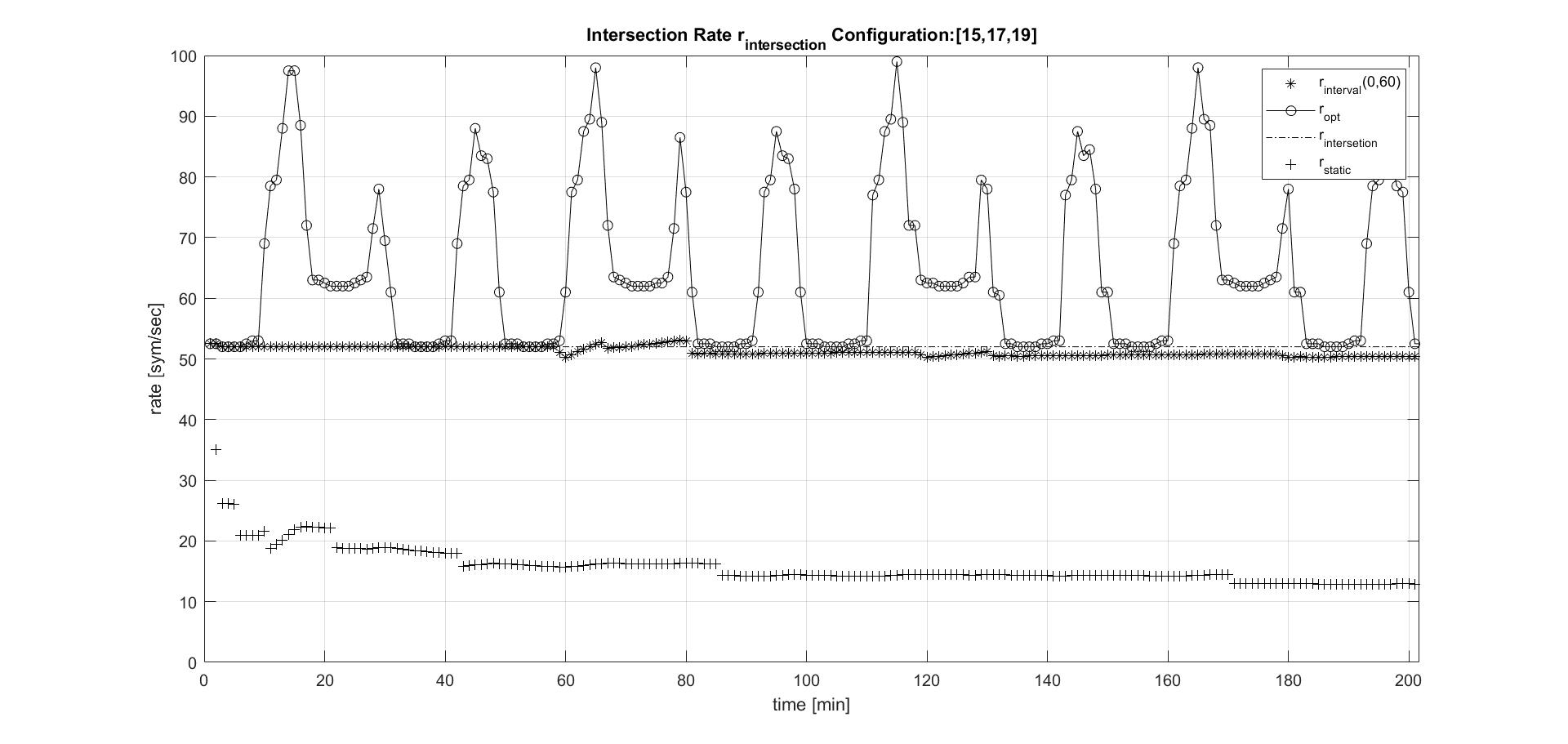}}%
\end{minipage}
\end{figure}

\begin{defn}
\emph{Network static rate} $r_{static}(\tau)$ with respect to the
static configuration $\varepsilon=\{\epsilon_{1},\epsilon_{2},...,\epsilon_{\tau}\}$
is the network overall throughput\footnote{since this is a \emph{linear multicast} framework, the throughput
of all designated sinks is the same, hence w.l.o.g. we have chosen
to focus our discussion from a single-sink perspective. } assuming all path variations are considered as static links at network
initialization (i.e, at $\tau=0$), hence increasing the size of the
field $F$. $r_{static}(\tau)=\frac{\stackrel[\xi=1]{\tau}{\sum}h(\xi)}{log(\hat{|F(\tau)|})}[symbols/sec]$
where $h(\xi)$ is the max-flow of the network for configuration $\epsilon_{\xi}$,
and $|\hat{F(\tau)}|>|F|\cdot\tau$.
\begin{defn}
\emph{Network intersection rate} $r_{intersection}(\tau_{1},\tau_{2})$
is the overall network throughput when using the intersection network
$\stackrel[t=\tau_{1}]{\tau_{2}}{\cap}G(t)$, inducing $r_{intersection}(\tau_{1},\tau_{2})=\frac{h_{int}(\tau_{1},\tau_{2})}{log|F|}$,
where $h_{int}(\tau_{1},\tau_{2})$ denotes the max-flow of $\stackrel[t=\tau_{1}]{\tau_{2}}{\cap}G(t)$.
\end{defn}
\end{defn}
When examining $r_{intersection}(0,\tau)$ in the case of our example
network, one can identify in Figure \ref{fig:techniques-comparision.-it}
that the rate stabilizes after a relatively short period of time.
This implies that a significant portion of the network is practically
fixed. Denote this ``stable'' max-flow value by $h_{int}^{stable}$
corresponds to a ``stable'' rate $r_{intersection}^{stable}$ .
\begin{cor}
\label{cor:expanding paths}In the example Iridium-based network,
there exists a set of $h_{int}^{stable}$ paths denoted by $P$, where
$p_{1},\ldots,p_{h_{int}^{stable}}\in P$ are fixed under network
dynamics. Note: this set is not necessarily expandable such that the
addition of $p_{h_{int}^{stable}+1},\ldots p_{h(\tau)}$ will achieve
the network\textquoteright s max-flow at time instance $\tau$.
\end{cor}
The next example is aimed at demonstrating \ref{cor:expanding paths}
and show that the addition of $p_{h_{int}^{stable}+1},\ldots p_{h(\tau)}$
will not achieve the network\textquoteright s max-flow at time instance
$\tau$.
\begin{example}
Let us examine the time-varying butterfly network depicted in Figure
\ref{fig:Time-variant-butterfly-network}. At $t=1$ (Figure \ref{fig:time variant butt in t=00003D1})
links $\#4,5$ are missing from the graph (but do appear at $t=2$).
Thus, the only path available to approach sink node $\#6$ is $P_{1}^{1}=\{1\rightarrow3\rightarrow7\rightarrow8\}$
(the numbering refer to the \textbf{edges}) while the paths to sink
$\#7$ achieving a max-flow (of 2) are $P_{1}^{2}=\{2\rightarrow6\},\ P_{2}^{2}=\{1\rightarrow3\rightarrow7\rightarrow9\}$
achieving $h_{6}=1,$and $h_{7}=2$. It is easy to see that at $t=2$,
Figure \ref{fig:time variant butt in t=00003D2},the achievable max-flow
is 2 for both sinks, i.e. $h_{6}=$$h_{7}=2$ is possible. In order
to achieve max-flow, edge $\#1$ and edge $\#7$ must be in \textbf{different}
paths (otherwise employment of path $\{1\rightarrow5\}$ is impossible,
essential for achieving max-flow). Thus, changing $P_{1}^{1}$ is
a necessity.
\begin{figure}[H]
\caption{\label{fig:Time-variant-butterfly-network}Time-varying butterfly
network}

\begin{minipage}[c][0.25\textheight]{0.45\columnwidth}%
\begin{center}
\subfloat[\label{fig:time variant butt in t=00003D1}At $t=1$]{

\includegraphics[height=0.15\textheight]{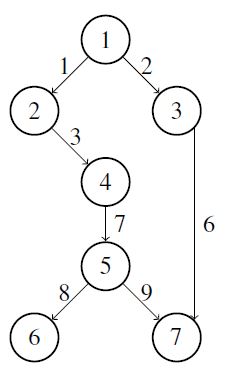}

}
\par\end{center}%
\end{minipage}\hfill{}%
\begin{minipage}[c][0.25\textheight]{0.45\columnwidth}%
\begin{center}
\subfloat[\label{fig:time variant butt in t=00003D2}At $t=2$]{

\includegraphics[height=0.15\textheight]{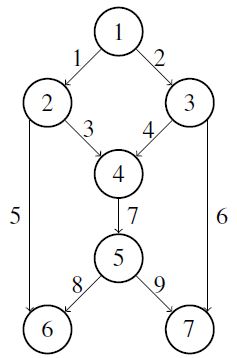}}
\par\end{center}%
\end{minipage}
\end{figure}
\end{example}

\subsubsection{\emph{Network Interval Rate}}
\begin{defn}
\emph{\label{def:Networks-interval-rate}} \emph{Network interval
rate} $r_{interval}(\tau,T)$ is the (average) network throughput
in interval time $\tau$ assuming the network code parameters must
be re-distributed over the entire network once every sub-interval
time $T$, $\tau<T$. The base rate within the $k$'th sub-interval
is the network intersection rate $r_{intersection}(\tau_{1},\tau_{2})$
for $\tau_{1}=(k-1)\cdot T,\;\tau_{2}=kT.$ The overall rate is defined
as the average on the entire interval $T$ (including the distribution
time):$r_{interval}(\tau,T)=\frac{\stackrel[k=1]{\tau/T}{\sum}r_{intersection}((k-1)\cdot T,kT)\cdot(1-\frac{t_{distribution}}{T})}{\tau/T}$$=\stackrel[k=1]{\tau/T}{\sum}r_{intersection}((k-1)\cdot T,kT)\cdot(\frac{T-t_{distribution}}{\tau})$

where $t_{distribution}$ is the time required for re-distributing
the tailored code-parameters throughout the entire network (see Equation
\ref{eq:t_distribution}).

\begin{figure}[H]
\caption{\label{fig:Interval-Rate-performance-as}Interval-Rate performance
as a function of the sub-interval time $T$; single source, sink nodes
are \{6, 13,15\}. Optimum rate is achieved for $T=10[min]$.}

\centerline{

\includegraphics[height=0.15\textheight]{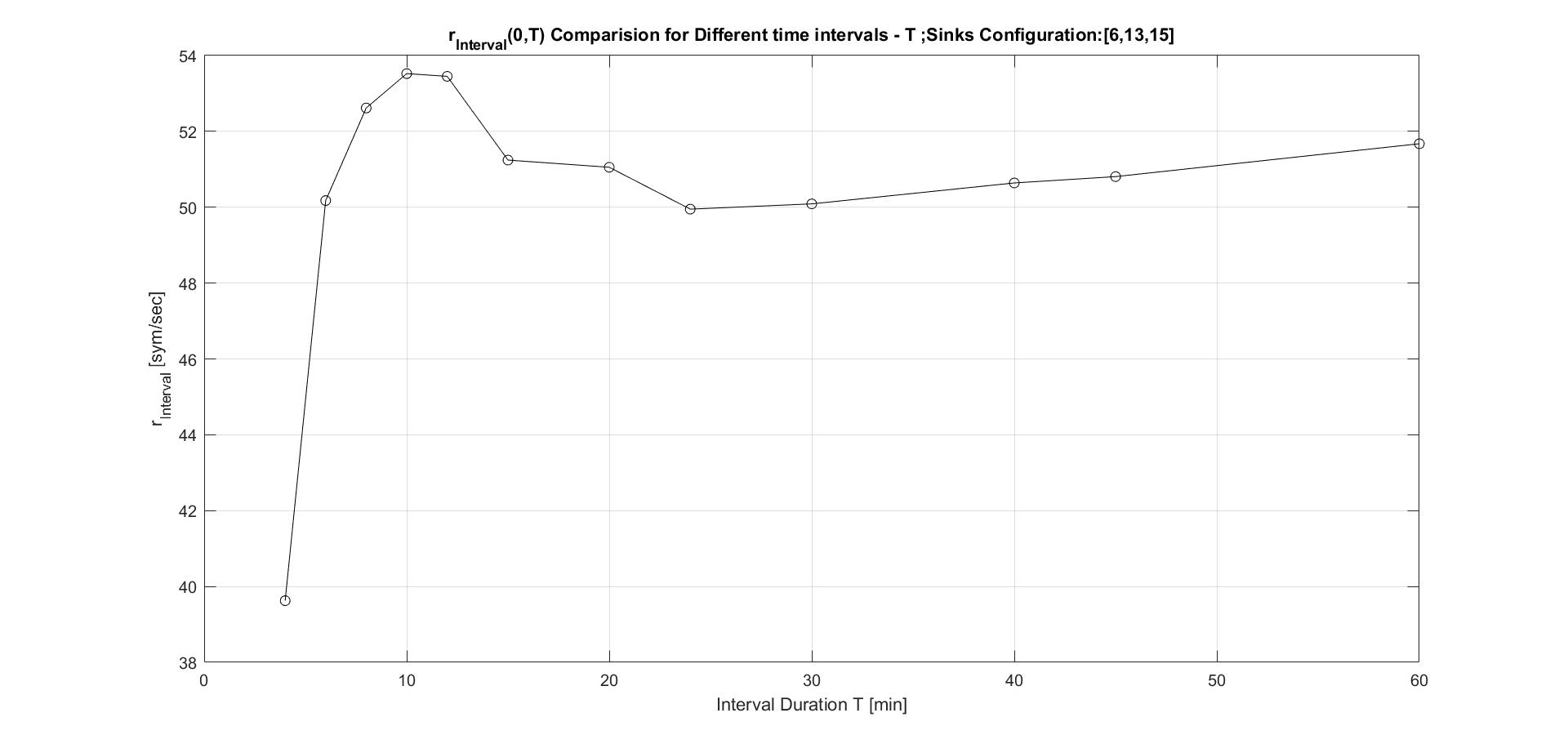}}
\end{figure}
\end{defn}
As demonstrated in Figure \ref{fig:Interval-Rate-performance-as},
there exists a certain \emph{sub-interval time} $T$, achieving the
optimum Network interval rate. In fact, in many of the examined configurations
such optimum exists. In addition, we note that for very long sub-intervals,
the network interval rate almost coincides with the network intersection
rate.
\begin{defn}
Let $t_{period}$ be the period time of $r_{opt}$, meaning $t_{perod}=t_{peak_{\#n+1}}-t_{peak_{\#n}}$.
For example, Figure \ref{fig:optimal-rate-bound} shows a peak in
$t_{peak_{\#1}}=30m$, and $t_{peak_{\#2}}=80m$ resulting $t_{period}=50m$. 
\begin{defn}
Let $\tau_{stable}$ denote the time achieving a stable \emph{intersection
rate} $r_{intersection}(\text{\ensuremath{t_{0}}},\tau)$ , with $t_{0}$
being the simulation's start time. 
\begin{align*}
\tau_{stable} & =min\{\tau|r_{intersection}(\tau,\infty)>precentile90(r_{intersection}(0,\infty))\}.
\end{align*}
 It is important to set this criterion and not arbitrarily choose
$t_{period}$ since in some of the cases there is a strong dependency
on the initialization conditions (the choice of $t_{0}$) such that
the upper bound for relaxation is $2\cdotp\tau_{period}$.
\begin{defn}
Let $T_{opt}$ denote the \emph{(optimal) sub-interval time} $T$
that results with the best achievable \emph{stable} rate $r_{interval}(\tau,T)$
for different $T's$. $T_{opt}=\underset{\tau_{stable}\leq T\leq\tau}{max}\{r_{interval}(\tau,T)|$$\tau>\tau_{stable}\}$
i.e, comparing different $T$'s after $\tau_{stable}$.\ref{fig:Interval-Rate-performance-as}
shows the comparission of different $T's$ (rates are after $\tau_{stable}$),
we can clearly see that $T_{opt}=10m$ in the example herein.
\end{defn}
\end{defn}
\end{defn}

\subsubsection{Techniques Comparison}

In order to identify which is the best network coding technique, we
shall compare the following rates: $r_{intersection}$, $r_{interval}(T_{opt})$,
and $r_{opt}$ ( $r_{opt}$ being the upper bound demonstrated in
Figure \ref{fig:optimal-rate-bound}).

It is interesting to note that for the majority of the studied topologies
and sink configurations, optimal value $T_{opt}$ exists, namely $r_{interval}(\tau,T_{opt})$
outperforms all the other network coding techniques. For all the other
studied topologies, no specific $T_{opt}$ value was found, and $r_{interval}$
monotonically increases with the sub-interval length $T$ such that
$r_{interval}(\tau,T_{opt})\rightarrow r_{intersection}(\tau,\infty)$.

\subsection{On the Existence of $T_{opt}$}

\begin{figure*}
\caption{\label{fig:Examining-Time-and}Examining Time and Frequency Domain
Characteristics of Both Topologies }

\includegraphics[width=0.5\columnwidth]{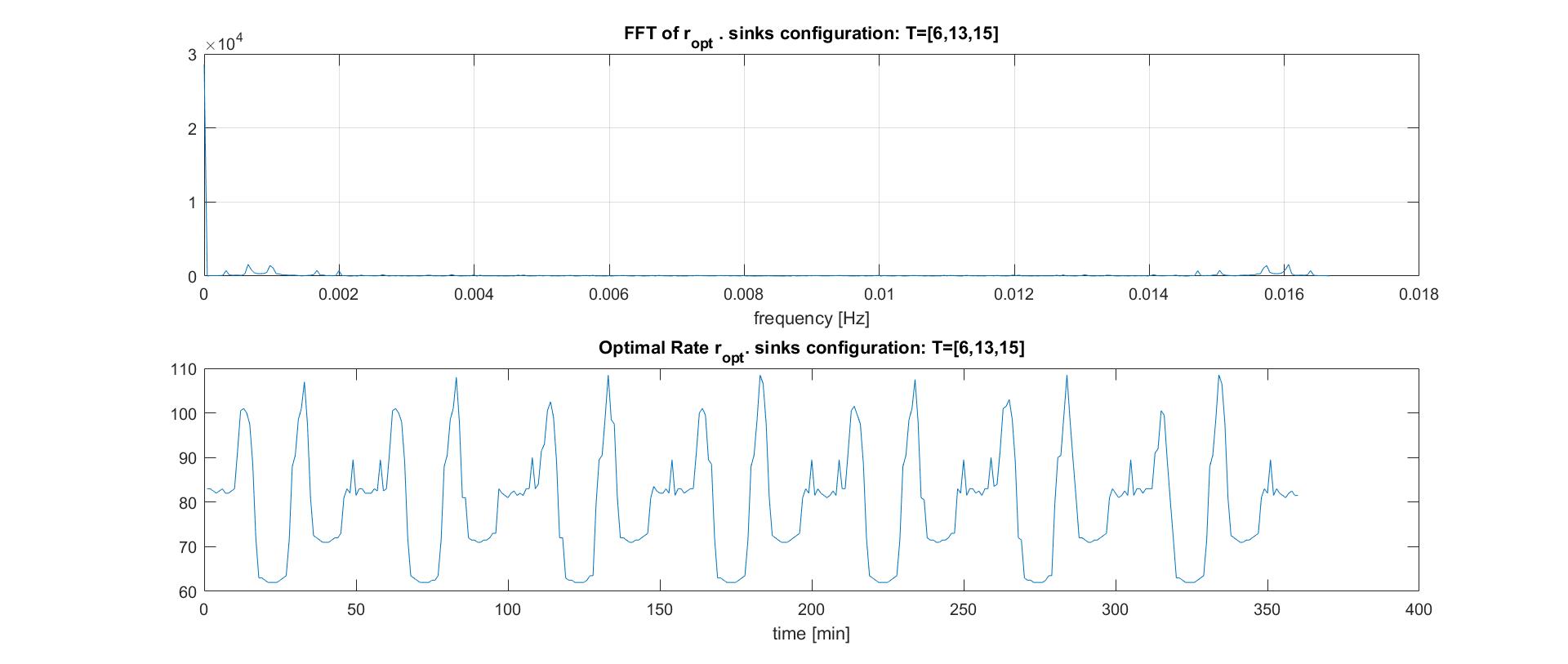}\includegraphics[width=0.5\columnwidth]{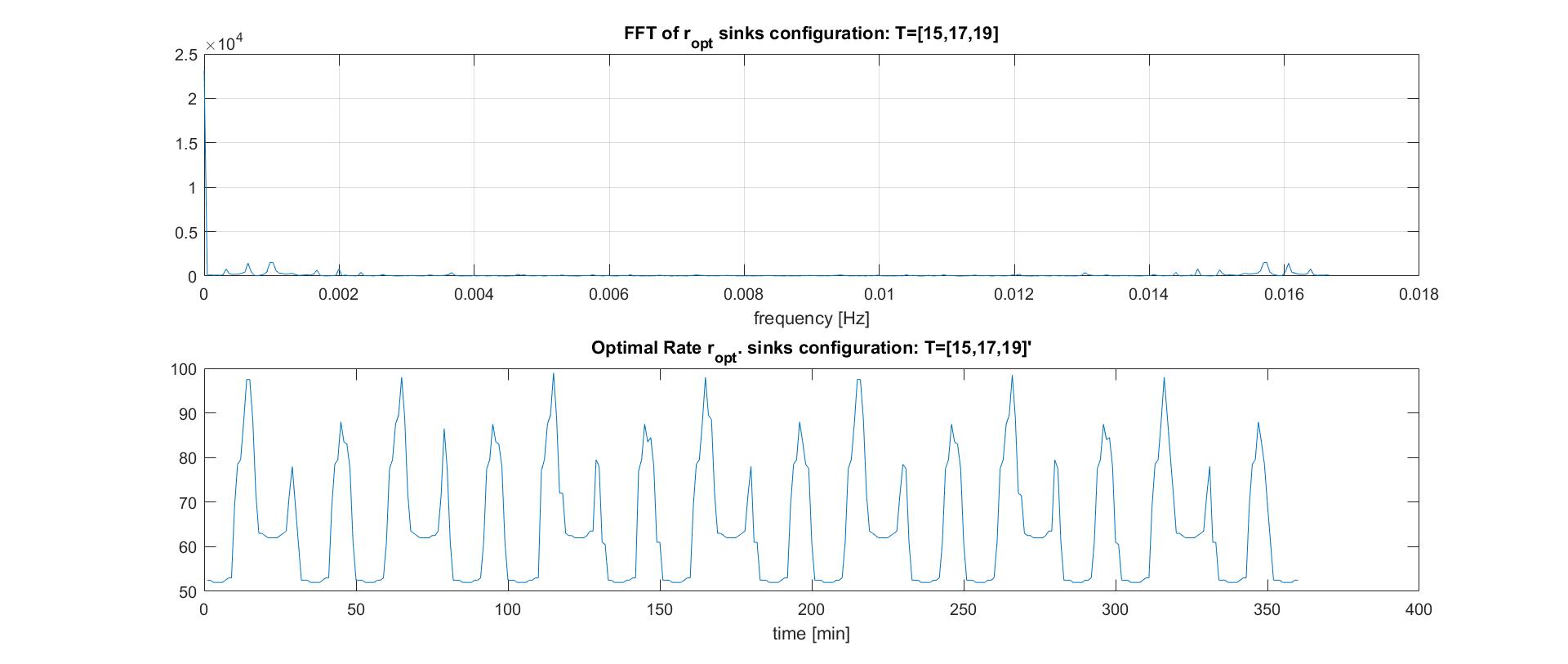}
\end{figure*}

Since the \emph{interval rate} was identified as the best network
coding scheme if specific $T_{opt}$ exists, it certainly is desirable
to determine whether such $T_{opt}$ exists without having to explicitly
design the network code. In this section we outline criteria for addressing
this problem.
\begin{example}
Let us compare two network configurations: one with the set of sinks
$T=\{6,13,15\}$ and the other with sinks $T=\{15,17,19\}$ demonstrated
in Figure \ref{fig:-for-different r interv}.

\begin{figure}
\caption{\label{fig:-for-different r interv}$r_{interval}$ for different
sink sets $T$ in both configurations}

\includegraphics[width=0.5\columnwidth]{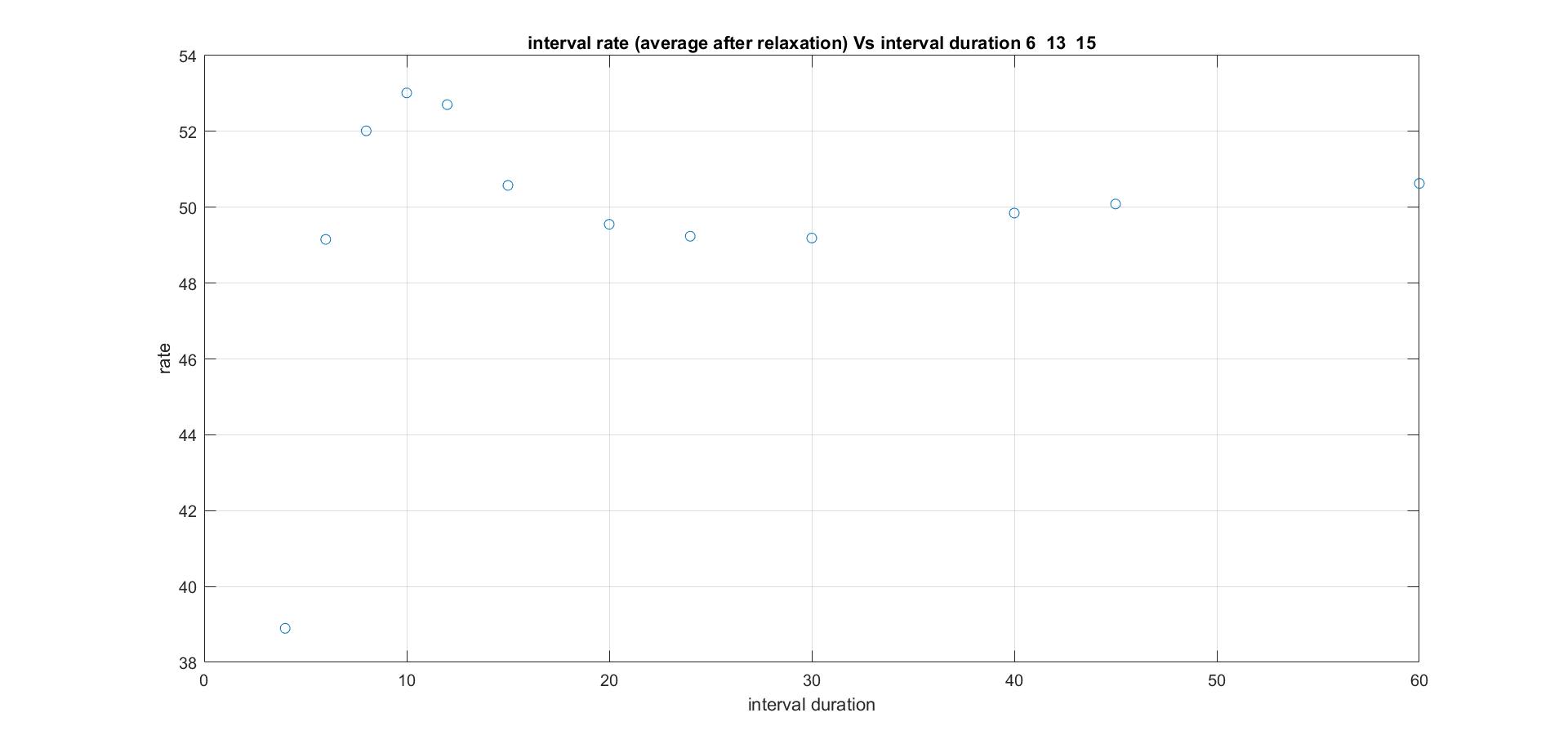}\includegraphics[width=0.5\columnwidth]{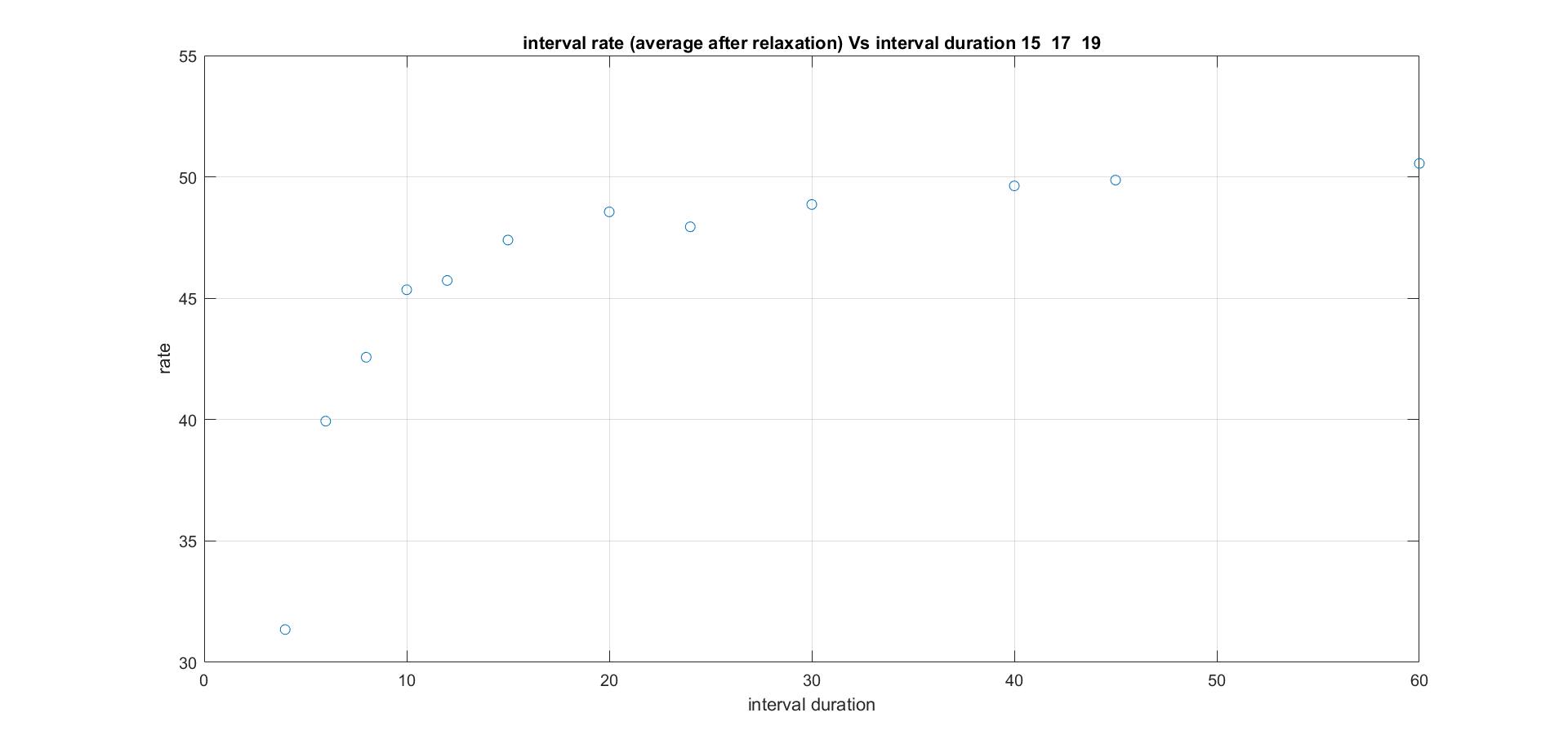}

\end{figure}
\end{example}
We clearly note that the configuration with $T=\{6,13,15\}$ has a
distinct optimum $r_{interval}$ for $T_{opt}=10[min]$, while the
rate in the configuration $T=\{15,17,19\}$ is monotonously increasing
such that $r_{interval}\rightarrow r_{intersection_{inf}}$. This
short example is very interesting because two quite similar topologies
(3 sinks, same source) need different coding scheme. One difference
between the configurations is that sinks $\{15,17,19\}$ are all in
the same orbital-plane, while$\{6\}$ and $\{13,15\}$are not. However,
other cross-plane sink configurations didn't necessarily resulted
the same outcome, meaning there is a more profound difference we must
indicate.We tried to build a model that is based on time and frequency
domain properties of the network instantaneous max-flow (i.e. $r_{opt}$)
which is a direct result of the network topology to find this criterion.
\begin{example}
Let us examine $r_{opt}$ of the two configurations as presented in
\ref{fig:Examining-Time-and}.
\end{example}

$r_{interval}(\tau,T)$ as defined in \ref{def:Networks-interval-rate}
does not take only the maximum value of $r_{opt}$, but is also influenced
by the $r_{opt}$ peak width, and number of peaks. Thus, examining
the signal also in the frequency domain (observing the Fourier transform
of $r_{opt}$) is desired. Unfortunately there was no correlation
between configurations that yields $T_{opt}$ and those that don't.

This forces us to look for different criteria, in the time domain,
but rather than employing only the maximum max-flow values, we shall
also consider different percentiles of $r_{opt}$ suggesting how long
does a certain topology is above a specific value:
\begin{enumerate}
\item \emph{PAPR }denotes the peak-to-average ratio of $r_{opt}(\tau)$
; $PAPR=\frac{max(r_{opt}(\tau))}{mean(r_{opt}(\tau))}$.
\item \emph{maxRa }denotes the maximum of  $r_{opt}(\tau)$ during the simulation;
$maxRa=\frac{\underset{\tau}{max}\{r_{opt}(\tau)\}}{r_{intersection}}$
\item \emph{rateR }denotes the maximum of  $r_{interval}(\tau,T_{opt})$
during the simulation; $rateR=\frac{max(r_{interval}(\tau,T))}{r_{intersection}}$\}
\item \emph{p50 }denotes the ratio $p50=\frac{max(r_{opt}(\tau))}{median(r_{opt}(\tau))}$\emph{
.}
\item \emph{p75 }denotes the ratio Median $p75=\frac{max(r_{opt}(\tau))}{precentile75(r_{opt}(\tau))}$
\end{enumerate}
other variations of percentiles were examined (80\%, 90\%, 25\%),
they didn't result different outcomes, thus weren't mentioned specifically. 

\emph{rateR }represents our ``unit under test'', $rateR<1$ values
represents configuration without $T_{opt}$. In order to compare the
different techniques, other criteria were also set to be unit-less.
\emph{PAPR }represents the ``classical'' criterion comparing peak
to average, while \emph{p50}, and \emph{p75}~increase the importance
of the rate distribution (i.e. testing for how long the signal stays
at higher rates). \emph{maxRa }emphasizes only the peak value of the
specific configuration. 

\begin{figure}[H]
\caption{\label{fig:Correlation-Matrix-between}Correlation Matrix between
Different Topology Criteria }

\centerline{

\includegraphics[height=0.22\textheight]{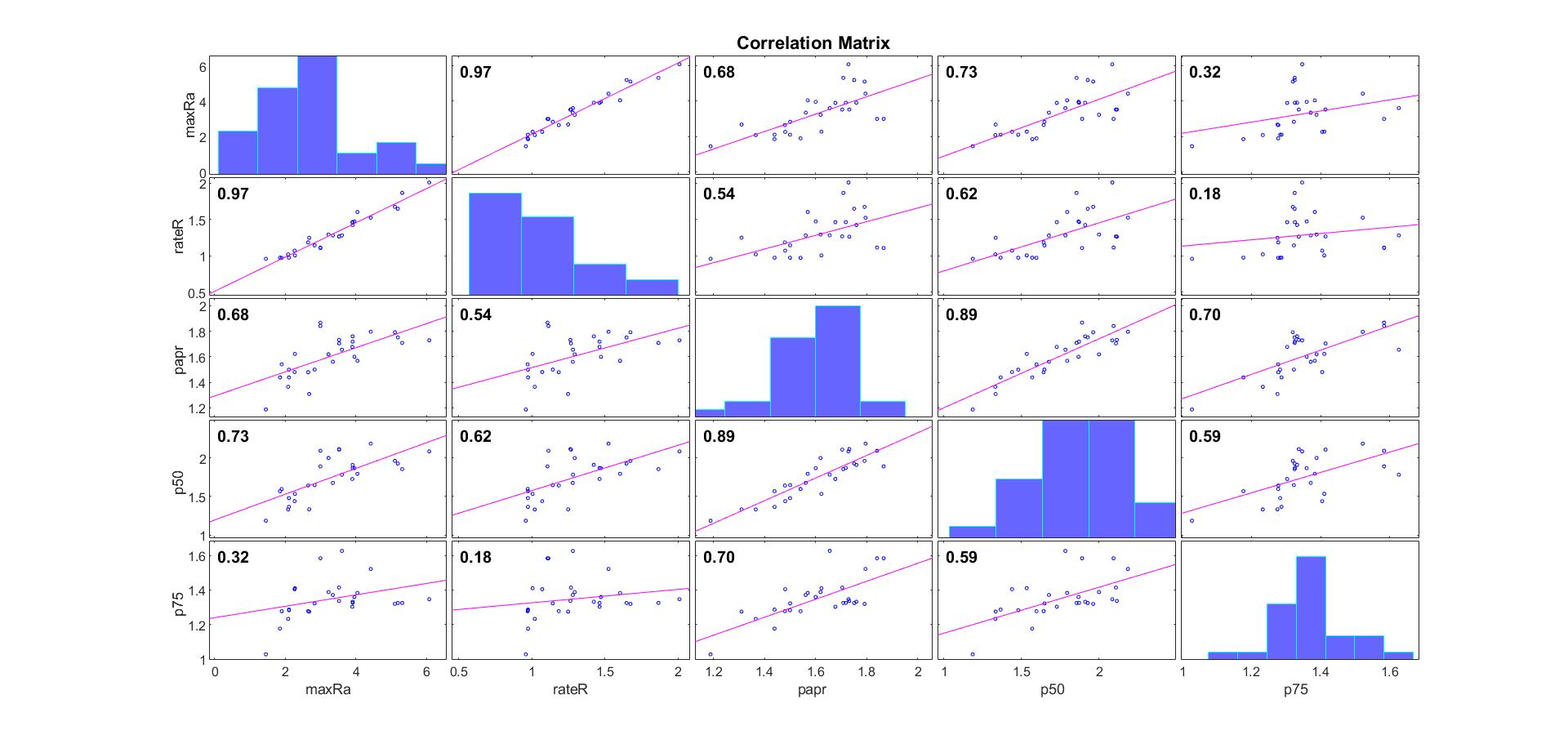}}
\end{figure}

Figure \ref{fig:Correlation-Matrix-between} shows a comparison between
the correlations of the different criteria presented for 29 different
configurations (varying number of sinks, number of orbital-planes).
As can be seen in Figure \ref{fig:Correlation-Matrix-between}, the
strongest correlation between \emph{rateR} to any other criterion
is to \emph{maxRa}, hence implying a connection, and a desire to test
for a specific threshold determining rather $T_{opt}$ exists. This
result supports the lack of correlation in the frequency domain analysis
since inhere too, the criterion chosen (\emph{maxRa) }does not express
any temporal aspect unlike p50 or p75. This is quite surprising since
$r_{interval}(\tau,T)$ by definition has a strong time dependency.

\subsubsection*{Setting the Threshold}

In Figure \ref{fig:Correlation-Matrix-between} we focus on the histogram
of the chosen criterion $maxRa=\frac{\underset{\tau}{max}\{r_{opt}(\tau)\}}{r_{intersection}}$
compared to $rateR=\frac{max(r_{interval}(\tau,T))}{r_{intersection}}$.
We can clearly see, that configurations with the ratio $maxRa<2$
are more likely to not have an optimum for $r_{interval}(0,T)$ (the
corresponding configurations have $rateR<1)$.

\begin{figure}[H]

\caption{\label{fig:Threshold-Examination}Correlation Examination to Set Threshold
for Different Configurations. black line: maxRa, blue line: rateR}

\centerline{\includegraphics[height=0.18\textheight]{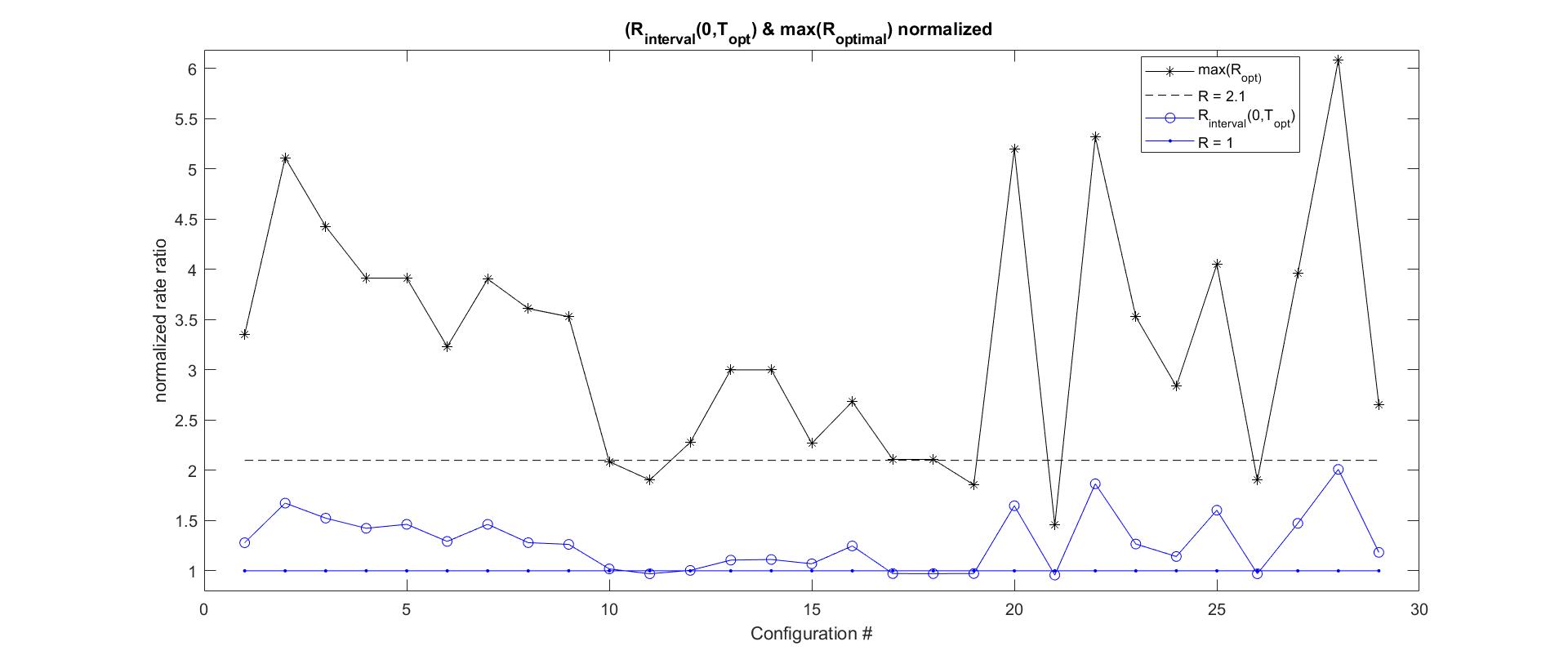}

}
\end{figure}

Figure \ref{fig:Threshold-Examination} shows that the correlation
presented in Figure \ref{fig:Correlation-Matrix-between}  indeed
indicates that a configuration with no distinct $r_{intersection}(0,T)$
maxima (i.e. with no $T_{opt},$resulting with $rateR<1)$ has a $\frac{max(R_{opt})}{R_{intersection}(0,\infty)}<2.1$
ratio.

To conclude this section, not only that better approaches than \emph{static
network coding} were presented for dealing with topology variations,
but an efficient way of finding the most appropriate coding scheme
was advocated. Determining whether a certain configuration has an
optimal $T_{opt}$ is the key, and it no longer requires computing
all different options of $r_{interval}(\tau,T)$, but only computing
$max(r_{opt})$, and $r_{intersection}(0,\infty)$. This suggested
test is based on 29 different sink configurations intended for spanning
all sorts of topological aspects of the network (sink numbers, cross
orbital-planes combinations). Nevertheless, the threshold for the
suggested test might change under new configurations, this without
decreasing the strength of this criterion.

\section{Conclusion}

A schematic implementation of network coding was presented for a swarm
of communicating satellites, conceptually based on the Iridium system.
Close-to-realistic network dynamics model and its parameters have
been derived from the Systems Tool Kit (STK) simulation environment.
In particular, we introduced the notion of \emph{generalized acyclic
network}, which promoted effective generation of \emph{linear network
codes} for what was considered until now to be a cyclic network.

One of the major challenges of network coding under realistic network
conditions is the dynamic changes that the network experiences over
time. To cope with this challenge, several methods were suggested
and compared, among these is the known \emph{static network coding
}approach that is traditionally used to tackle link failures. It was
demonstrated that in all cases, relating to our chosen example network,
static network codes under-performed compared to the methods presented.
Importantly, a simple test was formulated so as to be able to determine
which of the coding schemes is best to employ in a given scenario.
The test is based solely on the topology of the network, more explicitly
- the relation between the peak max flow of the network and the max
flow of the \emph{intersection network}.

\bibliographystyle{plain}
\bibliography{bibliographyDynamicSatellite}

\end{document}